\documentclass[11pt,a4paper,singlecolumn]{article}
\usepackage[utf8]{inputenc}

\usepackage{color,xcolor, url, float,  verbatim, changepage}
\usepackage{graphicx}
\usepackage[linesnumbered,vlined,ruled]{algorithm2e}
\usepackage{setspace}
\usepackage{array, multirow, multicol, tabu} 
\usepackage{mathtools,amsmath,amssymb} 
\usepackage{hyperref}
\usepackage{lmodern}
\usepackage[T1]{fontenc}
\usepackage{textcomp}
\usepackage{booktabs}
\usepackage{caption}
\usepackage{natbib} 
\newcommand{\ModelName}{JPoNG{ }}
\DeclareMathOperator{\sign}{sign}
\usepackage[margin = 26mm]{geometry}

\usepackage{amsthm}
\newtheorem{proposition}{Proposition}

\def\halmos{\mbox{\quad$\square$}}

\definecolor{mediumblue}{rgb}{0.0, 0.0, 0.8}

\usepackage{xcolor}
\usepackage{soul}

\newcommand{\cvar}{\text{CVaR}}
\newcommand{\Ne}{\mathcal{N}^{\text{e}}}
\newcommand{\Ng}{\mathcal{N}^{\text{g}}}
\newcommand{\Te}{\mathcal{T}^{\text{e}}}
\newcommand{\Tg}{\mathcal{T}^{\text{g}}}

\newcommand{\xe}{\boldsymbol{x}^{\text{e}}}
\newcommand{\xg}{\boldsymbol{x}^{\text{g}}}
\newcommand{\xeg}{\boldsymbol{x}}

\newcommand{\ye}{\boldsymbol{y}^{\text{e}}}
\newcommand{\yg}{\boldsymbol{y}^{\text{g}}}

\newcommand{\de}{\boldsymbol{d}^{\text{e}}}
\newcommand{\dg}{\boldsymbol{d}^{\text{g}}}

\newcommand{\Fv}{\boldsymbol{F}^{\nu}}

\newcommand{\CeO}{\boldsymbol{C}^{\text{e}}_1} 
\newcommand{\CeT}{\boldsymbol{C}^{\text{e}}_2} 
\newcommand{\CgO}{\boldsymbol{C}^{\text{g}}_1} 
\newcommand{\CgT}{\boldsymbol{C}^{\text{g}}_2} 

\newcommand{\AeO}{\boldsymbol{A}^{\text{e}}_1} 
\newcommand{\AeT}{\boldsymbol{A}^{\text{e}}_2} 
\newcommand{\AgO}{\boldsymbol{A}^{\text{g}}_1} 
\newcommand{\BeO}{\boldsymbol{B}^{\text{e}}_1} 
\newcommand{\BeT}{\boldsymbol{B}^{\text{e}}_2} 
\newcommand{\BgO}{\boldsymbol{B}^{\text{g}}_1} 

\newcommand{\Ae}{\boldsymbol{A}^{\text{e}}} 
\newcommand{\Ag}{\boldsymbol{A}^{\text{g}}} 
\newcommand{\Be}{\boldsymbol{B}^{\text{e}}} 
\newcommand{\Bg}{\boldsymbol{B}^{\text{g}}} 

\newcommand{\bT}{\boldsymbol{b}_2} 
\newcommand{\bb}{\boldsymbol{b}} 

\usepackage{authblk,stackengine}
\usepackage{colortbl}
\usepackage{arydshln}

\renewcommand{\arraystretch}{1.4}  
\newcommand\barbelow[1]{\stackunder[1.2pt]{$#1$}{\rule{.8ex}{.075ex}}}

\providecommand{\keywords}[1]
{
  \small	
  \textbf{\textit{Keywords---}} #1
}

\title{Power-Gas Infrastructure Planning under Weather-induced Supply and Demand Uncertainties}
\author{Rahman Khorramfar$^{1}$\footnote{Corresponding author: khorram@mit.edu}$^*$, Dharik Mallapragada$^{2}$, Saurabh Amin$^3$  \\
        \small $^{1}$ MIT Energy Initiative (MITEI) and Laboratory for Information $\&$ Decision Systems (LIDS), MIT \\
        \small $^{2}$ Chemical and Biomolecular Engineering Department, Tandon School of Engineering, New York University \\
        \small $^3$ Civil and Environmental Engineering (CEE) and LIDS, MIT}

\date{}

\begin{document}
\maketitle

\begin{abstract}
Implementing economy-wide decarbonization strategies based on decarbonizing the power grid via variable renewable energy (VRE) expansion and electrification of end-uses requires new approaches for energy infrastructure planning that consider, among other factors, weather-induced uncertainty in demand and VRE supply. An energy planning model that fails to account for these uncertainties can hinder the intended transition efforts to a low-carbon grid and increase the risk of supply shortage especially during extreme weather conditions. Here, we consider the generation and transmission expansion problem of joint power-gas infrastructure and operations planning under the uncertainty of both demand and renewable supply. We propose two distributionally robust optimization approaches based on moment (MDRO) and Wasserstein distance (WDRO) ambiguity sets to endogenize these uncertainties and account for the change in the underlying distribution of these parameters that is caused by the climate change, among other factors. 
Furthermore, our model considers the risk-aversion of the energy planners in the modeling framework via the conditional value-at-risk (CVaR) metric. 
An equivalent mixed-integer linear programming (MILP) reformulation of both modeling frameworks is presented, and a computationally efficient approximation scheme to obtain near-optimal solutions is proposed. 
We demonstrate the resulting DRO planning models and solution strategy via a New England case study under different levels of end-use electrification and decarbonization targets. Our experiments systematically explore different modeling aspects and compare the DRO models with stochastic programming (SP) results.
\end{abstract}


\section{Introduction}

Power and gas systems are two major energy vectors in modern societies and their long-term planning is the main enabler of the energy transition-- a global shift away from the current fossil fuel-based generation to a decarbonized one. However, planning for this transformation is becoming increasingly complex due to evolving generation and consumption patterns, along with factors related to climate change mitigation and adaptation. On one hand, variable renewable energy (VRE) is expected to largely replace fossil-based energy generation. While essential for decarbonization, their inherent intermittent nature poses significant challenges to the planning and operations of energy systems \citep{MoreiraEtal2021_ARO}. The uncertainty in VRE generation arises not only from short-term weather variability but also from long-term shifts caused by climate change \citep{StaffellPfenninger2018}.
On the other hand, the increasing electrification of end-uses, along with the growing deployment of alternative energy carriers such as biofuel and hydrogen, is fundamentally reshaping the nature of the coupling between gas and electricity infrastructure as the two main energy systems \citep{VonWaldEtal2022}.
Consequently, effective planning must account for these uncertainty sources and accurately represent the interdependency between different energy systems.

Capacity expansion models are widely used to assess infrastructure needs in response to evolving technologies, demand projections, weather variability, and policy scenarios. Among these, the generation and transmission expansion problem (GTEP) is a well-established framework that provides decision support for long-term energy infrastructure and operations decisions. {Defined from a central decision-maker's perspective, GTEP integrates the generation expansion problem (GEP) and the transmission expansion problem (TEP). In essence,} it determines the optimal sizing, location, generation types, and transmission lines under a set of constraints \citep{KoltsaklisEtal2018_GEP_review}. Most GTEPs in the literature adopt models with deterministic parameters and primarily focus on power systems planning, with only a few considering multi-vector infrastructure planning under uncertainty \citep{KlatzerEtal2022}. Despite their widespread use, deterministic models can under- or overestimate the required infrastructure, potentially leading to planning outcomes that are vulnerable to extreme weather outcomes.

Here, we consider a joint electric power and gas system, focusing on weather-induced uncertainty and its propagation on the uncertain parameters of supply (i.e., capacity factors for VREs) and demand. We develop two distributionally robust optimization (DRO) models to capture these uncertainties and ensure the robustness of the planning outcome toward various weather realizations and distributional shifts due to climate change. 
GTEP is usually carried out by different entities such as utility companies and system operators. Therefore, our proposed model includes a risk measure to account for the sensitivity of the decision-maker toward extreme parameter values. 

Developing approaches to capture parameter uncertainties has been the focus of mathematical programming from its early days \citep{Dantzig1955_SP}. These efforts have led to the emergence of several approaches, among which stochastic programming (SP), robust optimization (RO) and DRO are most notable. SP assumes a perfect characterization of the underlying distribution of uncertainty parameters, usually in the form of discrete scenario realizations. {Therefore, the objective of SPs is often time formulated as the expected value of a function that depend on possible realizations}. Many studies consider two-stage SP in the context of energy planning.
\citet{MunozEtal2016_EJOR} propose a two-stage stochastic GTEP under uncertainty of demand and supply potentials of solar, wind and hydropower. \citet{ZhanEtal2016} develop a multistage decision-dependent SP for GEP in which the demand is uncertain and the investment decisions affect the electricity prices. 

{Coordinated planning of power and gas systems is the focus of some studies that model uncertainty by SP.} A GTEP is proposed in \citep{ZhaoEtal2017_2SP_power_gas} under power-gas demand growth uncertainty. A similar approach is proposed in \citep{NunesEtal2018} with long-term price and demand growth uncertainties and short-term generation potential of VREs. 
Despite having mathematical properties that contribute to its performance, the SP has shortcomings that are particularly salient in the context of energy systems planning. First, a proper representation of a distribution may require a large scenario size, leading to the curse of dimensionality and limited tractability of the problem \citep{Moret2020_EJOR}. Second, it assumes a perfect knowledge of the underlying distribution whereas scenarios are usually based on limited historical records whose distribution can change, for example, due to climate change and energy regulations. Third, SP with discrete scenarios may lead to solutions that perform poorly on out-of-sample scenarios \citep{BertsimasEtal2023_MSOM}. In power-gas systems, this may amount to a lack of resiliency and robustness toward extreme scenarios, leading to severe resource inadequacy.  

Robust optimization partially alleviates SP's shortcomings by hedging the solution against the worst parameter realization within a specified \textit{uncertainty set}. RO can be utilized without strong assumptions about the underlying distribution from which uncertain parameters are drawn. A three-stage RO is proposed in \citep{MoreiraEtal2021_ARO} for a power system GTEP with renewable and climate uncertainties. In \citep{HeEtal2017RO_power_gas}, robust optimization is applied to a power-gas GTEP with wind power uncertainty. 
Two frequent critiques of RO are its over-conservatism and computational burden \citep{RoaldEtal2023_survey}. While decision outcomes based on the RO approach can prepare the energy infrastructure for the worst possibilities, it may result in over-investment in certain assets, thus yielding significantly higher system costs. 

A unifying approach between SP and RO is DRO which offers a way to reduce the conservatism of RO while retaining the tractability of SP \citep{BertsimasEtal2023_MSOM}. Since the true distribution of uncertain parameters can be unknown, DRO uses scenarios of random parameters and domain knowledge to limit the true distribution to a family of distributions determined by an \textit{ambiguity set}. 
Depending on data availability and application area, the ambiguity sets can be defined considering various factors including shared moment or discrepancy between distributions \citep{RahimianMehrotra2019_DRO_review}. Discrepancy-based or metric-based ambiguity sets contain distributions whose probabilistic distance from the nominal empirical distribution is confined to a given positive value. Moment-based approaches build an ambiguity set containing distributions that share a set of moments (e.g., mean, covariance) that satisfy certain properties. {While both approaches offer a flexible framework to model uncertainty using partial information, DRO problems with moment-based ambiguity sets are more popular due to computational advantage, interpretability (intuitive statistical measures), and its connection to risk measures \citep{PourahmadiKazempour2021_GEP_DRO}. DRO models with metric-based ambiguity sets such as Wasserstein distance have gained popularity in recent years as they offer finite-sample efficiency and strong theoretical foundations, and can admit tractable reformulations
\citep{MohajerinKuhn2018_MathProg}.} 

The modeling prowess that DRO offers has made it an appealing choice in recent years for various problems in energy systems, including capacity expansion problems (Table~\ref{tab:lit-review}), distribution system \citep{Zare2018_TPS}, smart city energy system design \citep{Li2023data_AE}, virtual power plant operations \citep{Cao2025_VPP}, critical energy infrastructure \citep{BelleEtal2023}, and multi-energy systems \citep{Son2024_DRO}.
Table~\ref{tab:lit-review} provides an overview of the DRO applications in the context of GEP, TEP, and GTEP. 
Several studies consider the generation expansion problem in power systems. \citet{PourahmadiEtal2019_DRO_CC} model the long-term uncertainty of demand growth via discrete scenarios and short-term uncertainty of wind production through a moment-based ambiguity set. The authors also consider a chance constraint (CC) to adjust the conservatism of solutions and enhance out-of-sample performance. In a subsequent paper, \citet{PourahmadiKazempour2021_GEP_DRO} include modality information in the ambiguity set definition and propose CC and conditional value-at-risk (CVaR) as risk measures and find that CVaR presents better out-of-sample performance in the case of insufficient data to accurately estimate moments. \citet{Guevara2020_AE_GEP} apply XGBoost, a variable selection tool popular in machine learning applications (ML), in the construction of the ambiguity set to reduce the number of uncertain parameters. A wind power production uncertainty under a moment-based ambiguity set is proposed in \citep{HuEtal2022}. 

{\citet{Alvarado2018_TPS_TEP} and \citet{Alvarado2022_TPS_TEP} consider TEP and model the failure probability of components (i.e., generators and transmission lines) as moment-based DRO.
For the same problem, \citet{Velloso2020_TPS_TEP} model the short-term net load uncertainty via DRO and long-term demand growth and VRE adoption by SP.}
\citet{ChenEtal2023_Wassestein} consider metric-based ambiguity set with Wasserstein distance and CC to model the wind power uncertainty. 
A few studies consider GTEP and use DRO to model uncertainty parameters in power system planning. \citet{HajebrahimiEtal2020_GTEP_DRO} focus on the rising demand for electricity due to transportation electrification and propose a moment-based ambiguity set to capture the uncertainty of demand and wind power generation. Demand side flexibility and concentrating solar power along with a penalty for VRE generation is considered in \citep{ChenEtal2022_DRO_GTEP}. \citet{XieEtal2023Energy} develop a bi-objective optimization model and consider a metric-based approach to model the demand and generation potential of VRE.  A GTEP for inter-regional power systems planning is proposed in {\citep{Kang2023_EJOR_GTEP} in which the inter-regional transmission expansion decisions are determined by the central planner of the leader's problem and generation expansion decisions are made within regions. The authors model the long-term parameter uncertainty by DRO with first moment, and short-term perturbation by RO with a budgeted box uncertainty set. }

Planning under uncertainty naturally raises the question of quantifying and managing the risk associated with the worst realization of uncertain parameters. Here, we consider a mean-CVaR objective function to reflect the risk aversion of the decision maker in avoiding high-impact and low-probability scenarios. {This objective further allows a trade-off between the expected outcomes and the risk measure}. CVaR is a convex and coherent risk measure, making it a common risk index choice in a wide range of applications including energy systems optimization \citep{RoaldEtal2023_survey}. The mean-CVaR offers a flexible framework for balancing between the expected value function (risk-neutral) and CVaR measure (risk-averse).

Despite the importance of modeling the interdependencies between energy vectors, there is no substantial study for the joint GTEP of power-gas systems, as evident from Table~\ref{tab:lit-review}. Moreover, most studies consider uncertainty in either supply or demand parameters. This paper bridges the gap in the literature by offering both methodological and practical contributions. \textbf{First}, we present GTEP with CVaR measure for {co-optimized power-gas planning as two distributionally robust optimization models to jointly model the uncertainty in demand and supply. The first model is constructed based on a moment-based ambiguity set (MDRO) and the second model based on Wasserstein distance (WDRO).} \textbf{Second}, in the MDRO, the proposed ambiguity set embeds spatial correlation between nodes without introducing new set of variables and constraints. \textbf{Third}, we developed an easily implementable approximate method to efficiently solve the resulting MILP to near optimality. Unlike available decomposition algorithms that usually perform successfully in either SP or DRO models, the proposed method is generic and can be applied to both SP and DRO models with minimal tailoring. 
\textbf{Fourth}, the framework is applied to the realistic power-gas network in the New England region under various heating electrification levels. {The experiments are designed to demonstrate the importance of uncertainty,  the impact of co-optimizing power-gas networks, the isolated impact of demand or supply uncertainty, the value of risk measure, and out-of-sample performance for each modeling paradigm. }

\begin{table}
\small 
\caption{Overview of the related studies as compared to this paper on key modeling aspects}
{\footnotesize
\setlength{\tabcolsep}{7pt}
\renewcommand{\arraystretch}{1.1} 
\begin{tabular}{c|llllllll}
\toprule
\multirow{2}{0.08\columnwidth}{Reference} &
\multirow{2}{0.05\columnwidth}{Problem class}& \multirow{2}{0.04\columnwidth}{Energy vector}& \multirow{2}{0.1\columnwidth}{Uncertain parameters}& \multirow{2}{0.05\columnwidth}{Solution approach}& \multirow{2}{0.07\columnwidth}{Ambiguity set}& \multirow{2}{0.05\columnwidth}{Risk measure}& \multirow{2}{0.12\columnwidth}{Granularity}\\ \\ \midrule \midrule 
\multirow{2}{0.13\columnwidth}{\citet{Alvarado2018_TPS_TEP}}& TEP & \multirow{2}{0.04\columnwidth}{Power} & \multirow{2}{0.12\columnwidth}{Component failure} &  \multirow{2}{0.07\columnwidth}{Benders} & \multirow{2}{0.05\columnwidth}{moment-based} & - & \multirow{2}{0.17\columnwidth}{30-minute, 1 year, 45~rep.~days}\\ \\  \midrule
\multirow{2}{0.13\columnwidth}{\citet{PourahmadiEtal2019_DRO_CC}}& GEP & \multirow{2}{0.04\columnwidth}{Power} & \multirow{2}{0.12\columnwidth}{Wind energy gen.} &  \multirow{2}{0.07\columnwidth}{EF as MISOCP$^\dagger$} & \multirow{2}{0.05\columnwidth}{moment-based} & CC & \multirow{2}{0.17\columnwidth}{hourly,~1~year, 10~rep.~days}\\ \\  \midrule
 \multirow{2}{0.13\columnwidth}{\citet{HajebrahimiEtal2020_GTEP_DRO}}& GTEP& Power&\multirow{2}{0.1\columnwidth}{Demand, VRE gen.}& \multirow{2}{0.07\columnwidth}{EF as MILP$^*$}  & \multirow{2}{0.06\columnwidth}{moment-based} & - & \multirow{2}{0.12\columnwidth}{hourly,~4~years, 1~rep.~day}\\ \\ \midrule
 \multirow{2}{0.13\columnwidth}{\citet{Velloso2020_TPS_TEP}}  & TEP & \multirow{2}{0.04\columnwidth}{Power} & \multirow{2}{0.12\columnwidth}{Demand/wind energy gen.} & \multirow{2}{0.08\columnwidth}{CCG$^\ddagger$} & \multirow{2}{0.06\columnwidth}{moment-based} & - & \multirow{2}{0.15\columnwidth}{4-hour, 1 years, 1 rep. day} \\ \\ \midrule
\multirow{2}{0.13\columnwidth}{\citet{Guevara2020_AE_GEP}} & GEP & \multirow{2}{0.04\columnwidth}{Power} & \multirow{2}{0.1\columnwidth}{Fuel cost, heat rate}& \multirow{2}{0.07\columnwidth}{Benders- like alg.}& \multirow{2}{0.05\columnwidth}{metric-based} & - & \multirow{2}{0.12\columnwidth}{monthly, singly year}\\ \\ \midrule
\multirow{2}{0.17\columnwidth}{\citet{PourahmadiKazempour2021_GEP_DRO}} &  GEP & \multirow{2}{0.04\columnwidth}{Power} & \multirow{2}{0.12\columnwidth}{Wind energy gen.} & \multirow{2}{0.07\columnwidth}{EF as MISOCP} & \multirow{2}{0.05\columnwidth}{moment-based} & \multirow{2}{0.05\columnwidth}{CC, CVaR}& \multirow{2}{0.17\columnwidth}{hourly,~1~year, 10~rep.~days} \\ \\ \midrule
\multirow{2}{0.1\columnwidth}{\citet{HuEtal2022}} &  GEP & \multirow{2}{0.04\columnwidth}{Power} & \multirow{2}{0.12\columnwidth}{Wind energy gen.} & \multirow{2}{0.07\columnwidth}{EF as MISOCP} & \multirow{2}{0.05\columnwidth}{moment-based} & - & \multirow{2}{0.17\columnwidth}{hourly,~1~year, 8~rep.~days} \\ \\ \midrule
\multirow{2}{0.1\columnwidth}{\citet{ChenEtal2022_DRO_GTEP}}  & GTEP & \multirow{2}{0.04\columnwidth}{Power} & \multirow{2}{0.12\columnwidth}{Demand/wind energy gen.} & \multirow{2}{0.08\columnwidth}{EF as MILP} & \multirow{2}{0.06\columnwidth}{moment-based} & - & \multirow{2}{0.15\columnwidth}{hourly, 5 years, 1 rep. day} \\ \\ \midrule
\multirow{2}{0.13\columnwidth}{\citet{Alvarado2022_TPS_TEP}}  & TEP & \multirow{2}{0.04\columnwidth}{Power} & \multirow{2}{0.12\columnwidth}{component failure} & \multirow{2}{0.08\columnwidth}{CCG} & \multirow{2}{0.06\columnwidth}{moment-based} & - & \multirow{2}{0.15\columnwidth}{30-minute, 1 years, 1 rep. hour} \\ \\ \midrule
\multirow{2}{0.1\columnwidth}{\citet{ChenEtal2023_Wassestein}} & GEP & \multirow{2}{0.04\columnwidth}{Power} & \multirow{2}{0.12\columnwidth}{Wind energy gen.} & \multirow{2}{0.08\columnwidth}{EF as MILP} & \multirow{2}{0.06\columnwidth}{metric-based} & CC & \multirow{2}{0.15\columnwidth}{hourly,~5~years, 1~rep.~day} \\ \\ \midrule
\multirow{2}{0.12\columnwidth}{\citet{XieEtal2023Energy}} &  GTEP & \multirow{2}{0.04\columnwidth}{Power} & \multirow{2}{0.12\columnwidth}{Demand, VRE gen.} & \multirow{2}{0.08\columnwidth}{EF as MILP} & \multirow{2}{0.06\columnwidth}{metric-based} & \multirow{2}{0.08\columnwidth}{short- fall} & \multirow{2}{0.17\columnwidth}{hourly,~1~year, 1~rep.~day} \\ \\ \midrule 
\multirow{2}{0.12\columnwidth}{\citet{Kang2023_EJOR_GTEP}} &  GTEP & \multirow{2}{0.04\columnwidth}{Power} & \multirow{2}{0.12\columnwidth}{Demand, VRE gen.} & \multirow{2}{0.08\columnwidth}{EF as MILP} & \multirow{2}{0.06\columnwidth}{moment-based} & \multirow{2}{0.08\columnwidth}{-} & \multirow{2}{0.17\columnwidth}{yearly,~10~year} \\ \\ \midrule 
\multirow{2}{0.08\columnwidth}{This paper} &  GTEP & \multirow{2}{0.055\columnwidth}{Power \& gas} & \multirow{2}{0.1\columnwidth}{Demand, VRE gen.} & \multirow{2}{0.07\columnwidth}{Approx. procedure} & \multirow{2}{0.05\columnwidth}{moment, metric} & CVaR & \multirow{2}{0.17\columnwidth}{hourly,~1~year, 15~rep.~days} \\ \\ 
\bottomrule
\end{tabular}}

\vspace{0.06cm}

$^*$ EF as MILP: extensive form reformulation solved as a large-scale mixed-integer linear program\\$^\dagger$ EF as MSCOP: extensive form reformulation solved as a large-scale second-order conic program\\$^\ddagger$ CCG: cut-and-column generation algorithm
\label{tab:lit-review}
\end{table}

Throughout the paper, we use boldface symbols to denote matrices and vectors and use normal font for other data types. The rest of the paper is organized as follows. The generic GTEP for the joint power-gas model is introduced in Section~\ref{sec:problem-setting}. Section~\ref{sec:DRO-model} formulates the problem as two DRO models and derives the equivalent extensive form mixed-integer linear program. The solution approach is presented in Section~\ref{sec:solution-method}. The details of the case study and the computational setup for the numerical experiments are provided in Section~\ref{sec:case-study-comp-setup}. Results, analyses, and managerial insights are presented in Section~\ref{sec:comp-experiments}. Finally, the paper is concluded and the future directions are laid out in Section~\ref{sec:conclusion}.

\section{Problem Setting} \label{sec:problem-setting}
We consider general joint power-gas planning problem in which the operations of these two energy systems are interdependent. The planning is carried out {for a future target year by which year the system has to satisfy the growing demand and satisfy decarbonization targets among other constraints}.  In our formulation, power system operates on hourly and gas on daily resolution, {both represented by their corresponding network}. Following the literature \citep{TeichgraeberBrandt2019}, the planner solves the model over a set of representative days. These days are selected and weighted in a way to resemble the operations of the system over the whole planning horizon. 
In our model, the operations between power and gas systems are interdependent because: i) gas-fired power plants {in each power node} draw gas {from its adjacent gas nodes}; ii) the decarbonization target necessitates limiting emissions of CO$_2$ across both systems. Here, the term \textit{gas} is used to refer to natural gas (NG) and any drop-in low-carbon fuel (low-carbon fuel) such as renewable natural gas that can be transferred through the gas pipelines and have the functionality as NG \citep{ColeEtal2021}. We also assume that demand and supply parameters are given in a time series format. For example, the capacity factor for solar is given per each power node, each planning hour, and each scenario.

Let $\xe$ and $\xg$ be all \textit{investment} decisions in power and gas systems that are made before the planning horizon (time-independent variables), respectively. For the power system, these variables can include establishing new plants, storage technologies and transmission lines, and/or decommissioning of existing plants. For gas system, $\xg$ can represent establishing or decommissioning pipelines. 
The \textit{operational} variables of power and gas systems are denoted by $\ye$ and $\yg$, respectively. These variables for the power system represent generation, state of storage, flow, load shedding, and CO$_2$ emissions. For the gas system, $\yg$ can include decisions such as flow, load shedding, gas import, and CO$_2$ emissions. Also let $\mathcal{V}$ denote the set of VRE types (e.g., solar, onshore wind, offshore wind). 
The assumption here is that the operational problem is feasible for any investment decisions. This is not a restrictive assumption as load-shedding variables can handle any unsatisfied demand. Assuming that all notations are in matrix form, the deterministic joint power-gas GTEP can be formulated as:
\begin{subequations}\label{model:det}
\begin{align}
\min \ & \CeO \xe + \CeT\ye+\CgO \xg + \CgT \yg \label{det:obj}\\
\text{s.t. }\ & \AeO\xe+\BeO\ye=\de \label{det:c1}\\
& \Fv \AeT\xe+\BeT\ye=\bT & \nu\in \mathcal{V}\label{det:c2} \\
&\AgO\xg+\BgO\yg=\dg \label{det:c3}\\
& \Ae\xe+\Be\ye+\Ag\xg+\Bg\yg=\bb \label{det:c4}\\
& \xe,\xg \in \mathbb{Z}^+ \times \mathbb{R}^+,\ye,\yg \in \mathbb{R}^+\label{det:c5}
\end{align}
\end{subequations}

The objective function \eqref{det:obj} minimizes the bulk system cost for both systems. The \textbf{first term} includes i) investment cost for new plants, storage technologies, and transmission lines; ii) fixed operating and maintenance (FOM) cost for plants, storage facilities, and transmission lines; and iii) decommissioning cost for the existing power plants.  The \textbf{second term} represents the power system's operational expenditures including i) variable operating and maintenance (VOM) cost; ii) start-up cost; iii) fuel cost for thermal plants operating on non-gas fuels; and iv) load shedding cost. The \textbf{third term} captures i) investment cost for new pipelines and storage facilities; ii) FOM cost for pipelines and storage facilities; and iii) decommissioning cost of existing pipelines. Finally, the \textbf{fourth term } includes operating expenditures in the gas system including fuel import and load shedding costs. 

The first set of constraints~\eqref{det:c1} represents all constraints in the power system that involve a power load $\de$. Examples include power balance and renewable portfolio standard constraints.
Constraints~\eqref{det:c2} encapsulate all constraints that involve capacity factors $\Fv$ for the renewable generation type $\nu \in \mathcal{V}$. These constraints limits the generation from VRE to their production potential determined by the number of generators and their capacity factors. 
Similar to power system, constraints~\eqref{det:c3} represent all constraints that involve gas demand $\dg$. Finally, constraints~\eqref{det:c4} captures all the remaining constraints including i) operating plants determined by existing, new and decommissioned plants, unit commitment, ramping, generation limit for thermal plants, network, storage, and resource availability constraints for power system; ii) operating pipelines, flow, fuel import, and storage for gas system; iii) coupling constraints between power and gas system including the flow of fuel to power system, and joint emissions limit constraints. The constraints~\eqref{det:c5} determines the type of variables. Per common models in the literature \citep{Khorramfar2025_CRS}, some investment variables are assumed to be integer and the rest of the variables are continuous. The integer variables can represent the number of thermal plants established/decommissioned, and the establishment of transmission lines and pipelines (binary).

\section{DRO Model} \label{sec:DRO-model}
Consider the uncertainty of power demand, capacity factors of VREs, and gas demand, and let probability distribution $\mathbb{P}$ with finite and discrete support that contains all the possible scenarios (i.e., realizations) $\Xi=\{\xi^1, \xi^2,\ldots, \xi^S \}$ for all $s\in \mathcal{S}=\{1,2,\ldots, S\}$. Our assumption on the finiteness of the supports has a practical bearing as the projections for future supply-demand parameters are carried out in a limited number of ways, usually using well-known climate models \citep{Khorramfar2025_CRS}. Furthermore, the reformulation of continuous support leads to a semi-infinite integer program whose solution is not possible with commercial solvers \citep{BasciftciEtal2023COR}.

For each scenario $s\in \mathcal{S}$, let $\de_s$, $\Fv_s$, and $\dg_s$ denote the power demand, capacity factors of VREs, and gas demand for scenario $s$. We define $\xeg=(\xe,\xg)$ and $\boldsymbol{\omega}_s =(\de_s,\dg_s,\Fv_s)$ to simplify the exposition.
Let $\mathcal{P}(\Xi)$ be the ambiguity set that includes all distributions and is assumed to contain the true distribution $\mathbb{P}$. Here, $\Xi$ is the set of parameters that specifies the choice of ambiguity set. The elements of this true distribution are induced by the random vector $\boldsymbol{\omega}=\{\boldsymbol{\omega}_s|s\in \mathcal{S}\}$ defined over $\Xi$ where each scenario has the probability of $\mathbb{P}(\xi^s)=\boldsymbol{p}_s$. Assuming that $\lambda$ is the adjusting parameter between expectation and risk terms, the distributionally robust analogous model of \eqref{model:det} is:

\begin{subequations}
\begin{align*}
\min_{x\in \mathcal{X}} \ & \CeO \xe + \CgO \xg + \lambda \sup_{\mathbb{P}\in \mathcal{P}(\Xi)} \{\mathbb{E}_{\mathbb{P}}[V(\xeg,\boldsymbol{\omega})]\}+ (1-\lambda)\sup_{\mathbb{P}\in \mathcal{P}(\Xi)}\{\cvar^\alpha_{\mathbb{P}}(V(\xeg,\boldsymbol{\omega})) \}, 
\end{align*}
\end{subequations}

For a feasible $x\in \mathcal{X}$ and scenario $s\in \mathcal{S}$, $V(\xeg,\boldsymbol{\omega}_s)$ is defined as:

\begin{subequations}\label{model:val_func}
\begin{align}
V(\xeg,\boldsymbol{\omega}_s)= \min \ & \CeT \ye_s+ \CgT\yg_s \\
\text{s.t. }\ & \BeO\ye_s=\de_s-\AeO\xe \label{val_func:c1} \\
& \BeT\ye_s=\bT-\Fv_s\AeT\xe & \nu \in \mathcal{V} \\
&\BgO\yg_s=\dg_s-\AgO\xg \\
& \Be\ye_s+\Bg\yg_s=\bb-\Ae\xe-\Ag\xg  \label{val_func:c4}
\end{align}
\end{subequations}
 
Note that $V(\xeg,\boldsymbol{\omega}_s)$ is an LP for a given investment decision. Due to convexity of $V(x,\omega)$ {the maximum point is attainable, thus replacing the supremum operator with maximum yields:}
\begin{align}\label{model:dro_orig}
\min_{x\in \mathcal{X}} \ & \CeO \xe + \CgO \xg + \lambda \max_{\mathbb{P}\in \mathcal{P}(\Xi)}\mathbb{E}_{\mathbb{P}}[V(\xeg,\boldsymbol{\omega})]+(1-\lambda)\max_{\mathbb{P}\in \mathcal{P}(\Xi)}\cvar^\alpha_{\mathbb{P}}(V(\xeg,\boldsymbol{\omega})),  
\end{align}

The risk measure function \cvar{} is defined as \citep{SarykalinEtal2008_CVaR_VaR}:
$$
\cvar^\alpha_{\mathbb{P}}(V(\xeg,\boldsymbol{\omega}_s)) =\min_{\eta\in \mathbb{R}} \{\eta +\frac{1}{1-\alpha}\mathbb{E}_{\mathbb{P}}[\max(V(\xeg,\boldsymbol{\omega}_s)-\eta,0)] \}, 
$$

In effect, \cvar{} quantifies the potential loss in the worst $(1-\alpha)\%$ of the uncertain parameters' realization (scenarios). The optimum value of CVaR is attained at the confidence level $\alpha \in [0,1)$ is referred to as \textit{value-at-risk (VaR)}.

\subsection{{Moment-based} Ambiguity Set}
The existing moment-based ambiguity sets in the literature can rely on first and second moments \citep{PourahmadiEtal2019_DRO_CC, HuEtal2022}, resulting in a MISOCP reformulation. However, this can diminish the scalability of the problem \citep{Shehadeh2022TS} and undermine their applicability. Furthermore, the impact of incorporating second-order information on the solution outcome can often be minimal \citep{BasciftciEtal2023COR}. Some studies seek a remedy, for example by assuming spatially uncorrelated uncertain parameters or solely relying on functions of the first-order information \citep{HajebrahimiEtal2020_GTEP_DRO, ChenEtal2022_DRO_GTEP}. Instead, we define the ambiguity set that incorporates first-order moment as well as the spatial correlation between energy nodes { of the same vector (i.e., power or gas).} 

Let $\Ne$ and $\Ng$ be sets of nodes in power and gas systems, and $\mathcal{N}=\Ne \cup \Ng $ be set of all nodes. Let $\mathfrak{R}$ be the set of representative planning days, and $\mathfrak{T}_\tau$ be the set of hours in day $\tau \in \mathfrak{R}$. The set of planning periods for power system is $\Te=\{ t| t\in \mathfrak{T}_\tau,  \tau \in \mathfrak{R} \}$ and for gas system is $\Tg\equiv \mathfrak{R}$. 

\begin{figure}
    \centering
    \includegraphics[width=0.7\linewidth]{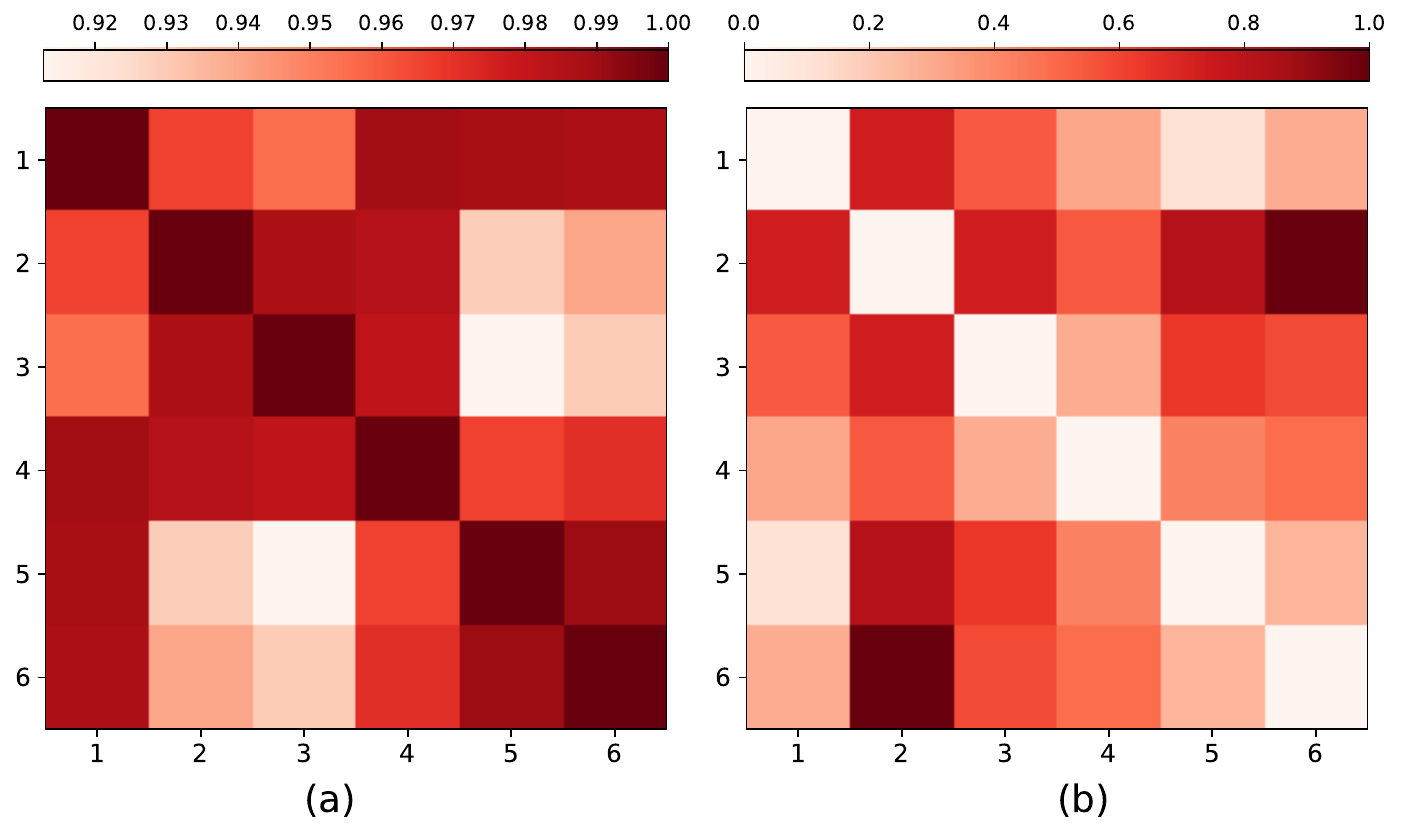}
    \caption{Correlation and distance between nodes. (a) average correlation between power nodes for demand time series across different scenarios; (b) normalized distances between power nodes. }
    \label{fig:corr_dist}
\end{figure}

{To motivate the deviation term in the definition of the ambiguity set, consider Fig.~\ref{fig:corr_dist} that shows the demand correlation and relative distance between power nodes for the case study. As shown, expectedly, closer nodes tend to present higher correlation because the weather variation usually impacts a large geographical area. For example, node 1 is closer to nodes 4, 5, and 6 for which the node has the highest correlation. Conversely, node 2, and nodes 5 and 6 are relatively apart; the correlation between node 2 and 5, and 2 and 6 is relatively weaker. Similar observations can be made for other stochastic parameters.}

Therefore, we incorporate the spatial correlation between energy node as part of the ambiguity set and introduce more notation. Let $\rho^\nu_{ij}$ be the average correlation across scenarios for capacity factor time series of VRE generator $\nu \in \mathcal{V}$ between power nodes $i$ and $j$. Similarly, let $\rho^{\text{e}}_{ij}\text{ and }  \rho^{\text{g}}_{kk'}$ be the average correlation between power nodes $i,j$  and gas nodes $k, k'$ across scenarios for power and gas demand time series, respectively. It is expected that closer nodes can have more similar profiles. Therefore, we further define $\ell_{ij}$ and $\ell_{kk'}$ as the inverse of the physical distance between power nodes $i$ and $j$, and gas nodes $k$ and $k'$, respectively. Then define the deviations terms as
\begin{align*}
\barbelow{\Delta}^{\text{e}}_i = \min\{ \kappa\ell_{ij}\rho^{\text{e}}_{ij} \ |\  \rho^{\text{e}}_{ij}<0, j\in \Ne, j\neq i \} \qquad  i\in \Ne  \\
\bar{\Delta}^{\text{e}}_i = \max\{\kappa\ell_{ij}\rho^{\text{e}}_{ij} \ |\  \rho^{\text{e}}_{ij}\geq 0, j\in \Ne,  j\neq i \} \qquad  i\in  \Ne  \\
\end{align*}
where $\kappa\geq 0$ is the adjusting factor set by the user.
{These quantities indicate the deviations from the nominal moment.}
Analogous quantities for capacity factors and NG demand are defined.  
Let $\mu^{\text{e}}_{it}$ and $\mu^{\text{g}}_{it}$ be the statistical mean of power and gas demand across scenarios at node $i$ at time $t$. Also, let $\mu^{\nu}_{it}$ be the statistical mean of capacity factors for VREs across scenarios at node $i$ at time $t$. Finally, the ambiguity set is defined as follows:

\begin{align}\label{ambguity_set}
\mathcal{P}(\mathcal{S}, \kappa) = \Bigg\{ \boldsymbol{p}\in \mathbb{R}^{S}_+ \lvert \ & \  \sum_{s\in \mathcal{S}}p_s =1,\\
& \mu^{\text{e}}_{it}+\barbelow{\Delta}^e_{i} \leq \sum_{s\in \mathcal{S}}p_s d^{\text{e}}_{sit} \leq \mu^{\text{e}}_{it}+\bar{\Delta}^e_{i} &  i \in \Ne, t\in \Te, \notag\\
& \mu^\nu_{it}+\barbelow{\Delta}^\nu_{i}  \leq \sum_{s\in \mathcal{S}}p_s F^\nu_{sit} \leq \mu^\nu_{it}+\bar{\Delta}^\nu_{i} & \nu \in \mathcal{V},  i \in \Ne, t\in \Te, \notag \\
& \mu^{\text{g}}_{k\tau}+\barbelow{\Delta}^g_{k} \leq \sum_{s\in \mathcal{S}}p_s d^{\text{g}}_{s k\tau} \leq \mu^{\text{g}}_{k\tau}+\bar{\Delta}^g_{k} & k \in \Ng, \tau \in \Tg \notag  \Bigg\}
\end{align}

Instead of the second moment, this definition allows the influence of other nodes depending on their proximity and correlation. {The example in Appendix~\ref{app:example} clarifies the notation.}
The problem ~\eqref{model:dro_orig} under ambiguity set~\eqref{ambguity_set} is a nonlinear and nonconvex model due to the presence of the max-min terms. 
To obtain a single-level reformulation of the problem, we first establish some properties for the value function of the second stage, namely $V(\xeg,\boldsymbol{\omega}_s)$. All proofs are relegated to the Appendix~\ref{app:proofs}.

\begin{proposition} \label{prop1}
The function $V(\xeg,\boldsymbol{\omega}_s)$ is a convex piecewise linear function of any given $\xeg\in \mathcal{X},\boldsymbol{\omega}_s$ such that $s\in \mathcal{S}$. 
\end{proposition}

\begin{proposition}\label{prop2}
  The problem \eqref{model:dro_orig} with the ambiguity set defined in \eqref{ambguity_set}, can be reformulated as the following MILP:
\end{proposition}

  \begin{subequations}\label{model:DRO-EF}
\begin{align}
\min \ & \CeO \xe + \CgO \xg + \lambda \Psi^{\text{obj}} +\delta  + (1-\lambda)\left[\frac{\alpha \xi}{1-\alpha}+\frac{\Psi^{\text{obj}}}{1-\alpha}\right]\\
\text{s.t. }\ &\Psi^{\text{cntr}}_s+\xi \geq 0& s\in S\\
& \delta \geq \CeT \ye_s+ \CgT\yg_s- \Psi^{\text{cntr}}_s & s\in S\\
&\eqref{val_func:c1}-\eqref{val_func:c4}\\
& \xe,\xg \in \mathbb{Z}^+ \times  \mathbb{R}^+,\ye_s,\yg_s,\xi \in \mathbb{R}\\
& \boldsymbol{\beta}^1, \boldsymbol{\beta}^2,\boldsymbol{\gamma}^1, \boldsymbol{\gamma}^2, \boldsymbol{\pi}^1,\boldsymbol{\pi}^2 \in \mathbb{R}^+
\end{align}
\end{subequations}

where

\begin{align*}
\Psi^{\text{cntr}}_{s} =&  \sum_{i\in \mathcal{N}^{\text{e}}}\sum_{ t \in \mathcal{T}^{\text{e}}} \left [\beta^1_{it}d^{\text{e}}_{sit} -\beta^2_{it}d^{\text{e}}_{sit} \right] +\sum_{i\in \mathcal{N}^{\text{e}}}\sum_{ t \in \mathcal{T}^{\text{e}}}\sum_{v\in \mathcal{V}} \left [\gamma^1_{vit}d^{\text{v}}_{sit} -\gamma^2_{vit}d^{\text{v}}_{sit} \right] + \sum_{k\in \mathcal{N}^{\text{g}}} \sum_{ \tau \in \mathcal{T}^{\text{g}}} \left [\pi^1_{k\tau} d^{\text{g}}_{sk\tau} -\pi^2_{vk\tau} d^{\text{g}}_{sk\tau} \right], s\in S \notag \\
\Psi^{\text{obj}}= & \sum_{i\in \mathcal{N}^{\text{e}}, t \in \mathcal{T}^{\text{e}}}\left[ (\mu^{\text{e}}_{it}+\bar{\Delta}^e_i)\beta^1_{it}-(\mu^{\text{e}}_{it}+\barbelow{\Delta}^e_i)\beta^2_{it} 
+\sum_{v\in \mathcal{V}}( (\mu^{v}_{it}+\bar{\Delta}^v_i)\gamma^1_{vit}-(\mu^{v}_{it}+\barbelow{\Delta}^v_i)\gamma^2_{vit}) \right] \\
&+\sum_{k\in \mathcal{N}^{\text{g}}, \tau \in \mathcal{T}^{\text{g}}}\left[ (\mu^{\text{g}}_{k\tau}+\bar{\Delta}^{\text{g}}_{k})\pi^1_{k\tau}-(\mu^{\text{g}}_{k\tau}+\barbelow{\Delta}^{\text{g}}_{k})\pi^2_{k\tau} \right]
\end{align*}

\subsection{{Wasserstein Ambiguity Set}}
{In order to define the Wasserstein distance, suppose that $\{{\omega}_1, {\omega}_2, \ldots, {\omega}_{|\mathcal{M}|}\}$ are support points on the sample space $\mathcal{S}$ with probability mass function $\mathbb{P}=\sum_{i\in \mathcal{M}} p_i\delta{\omega_i}$ where $\delta{\omega_i}$ is the Dirac measure of value 1 at parameter realization $\omega_i$ and 0 otherwise. Similarly let $\{{\zeta}_1, {\zeta}_2, \ldots, {\zeta}_{|\mathcal{K}|}\}$ be another sample set with probabilities $\mathbb{Q}=\sum_{j\in \mathcal{K}} q_j\sigma_{\zeta_j}$. If $\boldsymbol{\pi}$ is defined as the joint probability distribution over all available sample space (i.e., $\boldsymbol{\omega}\times\boldsymbol{\zeta}$), then $\mathbb{P}$ and $\mathbb{Q}$ become the marginal distributions of $\boldsymbol{\pi}$ and the type-L \textit{Wasserstein distance} between the two marginal distributions \citep{MohajerinKuhn2018_MathProg, book-SunConejo2021} is defined as:}

\begin{subequations}
    \begin{align}
    W_L(\mathbb{P},\mathbb{Q}) = &\min_{\pi\in \mathbb{R}^+} \sum_{j\in \mathcal{K}}\sum_{i\in \mathcal{M}} ||\omega_i-\zeta_j||^L \pi_{ij}\\
    &\sum_{j \in \mathcal{K}} \pi_{ij}=p_i & i\in \mathcal{M} \label{wass-marg1}\\
    &\sum_{i\in \mathcal{M}} \pi_{ij}=q_j & j\in \mathcal{K} \label{wass-marg2}
\end{align}
\end{subequations}

\noindent{where $||\cdot||^L$ is the L norm in the sample space domain. In other words, $W_L(\mathbb{P},\mathbb{Q})$ finds $\boldsymbol{\pi}$ in the set of all joint distributions of $\boldsymbol{\omega}$ and $\boldsymbol{\zeta}$ supported on $\mathcal{S}\times \mathcal{S}$ with marginal distributions $\mathbb{P}$ and $\mathbb{Q}$ specified by constraints \eqref{wass-marg1} and \eqref{wass-marg2}. Based on this definition, the ambiguity set over the Wasserstein ball of radius $\mathfrak{D}$:}

\begin{align}\label{model:wass}
    \mathcal{P}(\mathbb{Q}, \mathfrak{D}) = \{\mathbb{P}\in \mathcal{P}(\mathcal{S})|W_L(\mathbb{P},\mathbb{Q})\leq \mathfrak{D}\}
\end{align}

\begin{proposition}\label{prop3}
  The problem \eqref{model:dro_orig} with the ambiguity set defined by the Wasserstein distance in \eqref{model:wass}, can be reformulated as the following MILP.
\end{proposition}

\begin{subequations}
    \begin{align}
        \min \ &\ \CeO \xe + \CgO \xg+\lambda (\mathfrak{D}\beta^1+\sum_{j\in \mathcal{K}} \beta^2_j q_j)+ \notag \\
        & (1-\lambda)\left[\frac{1}{1-\alpha}( \mathfrak{D}\beta^1+\sum_{j\in \mathcal{K}} \beta^2_j q_j) +\eta \right] \notag \\   
      \text{s.t.}\ & ||\omega_i-\zeta_j||^L \beta^1 + \beta^2_j  \geq V(x,\omega_i)& i\in \mathcal{M}, j\in \mathcal{K}\\
      & ||\omega_i-\zeta_j||^L \beta^1 + \beta^2_j  \geq V(x,\omega_i)-\eta& i\in \mathcal{M}, j\in \mathcal{K}\\
      & ||\omega_i-\zeta_j||^L \beta^1 + \beta^2_j  \geq  0 & i\in \mathcal{M}, j \in \mathcal{K}\\
      & \BeO\ye_i+\AeO\xe=\de_i & i\in \mathcal{M} \\
& \BeT\ye_i+\Fv_i\AeT\xe=\bT & \nu \in \mathcal{V},i\in \mathcal{M} \\
&\BgO\yg_i+\AgO\xg=\dg_s & i\in \mathcal{M}\\
& \Be\ye_i+\Bg\yg_i+\Ae\xe+\Ag\xg=\bb & i\in \mathcal{M} \\
      &\xe,\xg \in \mathbb{Z}^+ \times \mathbb{R}^+,\ye_i,\yg_i, \beta^1\in \mathbb{R}^+, \beta^2_i,\delta^1_j, \delta^2 \in \mathbb{R}
    \end{align}
\end{subequations}

\section{Solution Method} \label{sec:solution-method}
{As presented in the previous section, the resulting reformulations are large-scale MILPs whose solution might be intractable for practical instances. However, energy planning problems need an efficient solution method that allows a range of sensitivity analyses in a reasonable timeframe.} Here, we develop an approximation approach to compute near-optimal solutions of the MILP. This approximation method relied on sequentially solving a series of linear programs to construct a feasible solution, thus the sequential construction method (SCM). The SCM is based on the observation that relaxing certain constraints and variables can significantly increase traceability while retaining key modeling features. Specifically for the problem considered here, relaxing the network constraints, relaxing the integrality of integer variables, and/or setting the integer variables to given values yields a model with significantly less computational burden. The integer variables in our model include i) $x^{\text{op}}_{ni}\in \mathbb{Z}^+, n\in\mathcal{N}, i\in \mathcal{H}$: number of operational thermal power plant type $i$ at node $n$ ($\mathcal{H}$ is set of thermal plant types); ii) $z^{\text{eInv}}_\ell \in \mathbb{B}$: if candidate transmission $\ell$ is established; iii) $z^{\text{gOp}}_{\ell}\in \mathbb{B}$: if pipeline $\ell$ is operational. {Let \textit{MILP}$^{\text{ref}}$ refer to an extensive form of any reformulation of problem~\eqref{model:det} under DRO or SP.} The SCM has the following steps:

\begin{enumerate}
\item Solve {\textit{MILP}$^{\text{ref}}$} with the following simplifications and get the solution $\bar{X}$:
\begin{itemize}
\item Remove all network constraints for the power system. It is equivalent to the \textit{copper-plate model} in which all nodes are assumed to be connected with all other nodes with infinite capacity. 
\item Relax integrality of integer variables
\end{itemize}
\item Solve {\textit{MILP}$^{\text{ref}}$} with the following simplifications and constraints and get the solution $\hat{X}$:
\begin{itemize}
\item Relax integrality of integer variables
\item For thermal plants, set $x^{\text{op}}_{ni} = \lfloor(\bar{x}^{\text{op}}_{ni}) \rceil, n\in\mathcal{N}, i\in \mathcal{H}$, where $\lfloor x \rceil$ is the rounding operator that returns the nearest integer value to $x$, and $\bar{x}^{\text{op}}_{ni}\in \bar{X}$
\item For VRE plants, add constraints $x^{\text{op}}_{ni} = \lfloor\bar{x}^{\text{op}}_{ni} \rfloor, n\in\mathcal{N}, i\in \mathcal{V}$, where $\lfloor x \rfloor$ is the floor operator that returns the greatest integer less than or equal to $x$, and $\bar{x}^{\text{op}}_{ni}\in \bar{X}$
\item Add constraints $z^{\text{gOp}}_\ell=1$ if $\bar{z}^{\text{gOp}}_\ell \geq \epsilon_1$ where $\bar{z}^{\text{gOp}}_\ell\in \bar{X}$, and $z^{\text{gOp}}_\ell=0$, otherwise.  
\end{itemize}
\item Solve {\textit{MILP}$^{\text{ref}}$} with the following simplifications and constraints and get feasible solution ${{X}}^*$:
\begin{itemize}
\item Set $x^{\text{op}}_{ni} = \hat{x}^{\text{op}}_{ni}, n\in\mathcal{N}, i\in \mathcal{H}$, where $\hat{x}^{\text{op}}_{ni}\in \hat{X}$.
\item Set $z^{\text{gOp}}_\ell =  \hat{z}^{\text{gOp}}_\ell$.
\item Add constraints $z^{\text{eInv}}_\ell=1$ if $\hat{z}^{\text{eInv}}_\ell > \epsilon_2$ where $\hat{z}^{\text{eInv}}_\ell\in \hat{X}$ , and $z^{\text{eInv}}_\ell=0$, otherwise.
\end{itemize}
\end{enumerate}

The resulting solution $X^*$ is feasible as it satisfies the integrality of integer solutions, network constraints, and all other problem constraints. 
The SCM can be directly applied to the formulation of the problem under MDRO, WDRO and SP models. The choice of $\epsilon_1$  and $\epsilon_2$ depends on energy topology and network parameters. Here, $\epsilon_1=0.01 $ and $\epsilon_2=0.3$ throughout the experiments. The implementation of SCM for the full description of is presented in Appendix~\ref{SCM4SP} for the two-stage stochastic programming implementation of the problem.

\section{Case Study and Computational Setup}\label{sec:case-study-comp-setup}
We formulate the DRO model for the joint planning of electricity-gas infrastructure, adapting operational constraints from the previously published deterministic model, JPoNG \citep{KhorramfarEtal2024_AE}. JPoNG determines the cost-optimal planning decisions for power and gas systems under a detailed representation of both systems' constraints. It allows different planning resolutions for each system; the power system operates on an hourly basis whereas daily resolution is the operational granularity of the gas system.
In our adapted formulation, the first stage decisions include establishing new power plants and battery storage, decommissioning existing power plants, establishing new transmission lines and pipelines, and decommissioning existing pipelines. The operational variables for the electric power system include power generation, storage, power flow, load shedding, and emissions amount. The gas network's operational variables consist of gas flow, NG and low-carbon fuel import, load shedding, and emissions amount. 

Our formulation represents the main investment and operational constraints. For the power systems, it includes operational plants, power balance, generation limit, ramping, load shedding limit, direct current approximation of the network flow, storage limit, storage state-of-charge, and land-use availability for VRE (resource availability). For the gas system, the constraints include operational pipelines, demand and supply balance, linearized network flow, gas and low-carbon fuel import limit, and storage constraints. The operations of both systems are interconnected through two sets of constraints. The first coupling constraints ensure the flow of gas to gas-fired plants in the power system, and the second set limits the CO$_2$ emission incurred by consuming NG in both systems. Discrete decision variables in the model include number of thermal and nuclear plants installed (integer) and candidate pipeline and transmission line added (binary variables). The rest of the variables are continuous.

The objective function of the formulated DRO model is to minimize the investment and operational costs for both power and gas systems. The investment cost includes i) investment cost in new generation technologies, ii) network expansion (transmission lines and pipelines),  iii) decommissioning cost for the existing power plants and pipeline, and iv) storage cost. The operational costs include i) variable operating and maintenance costs (VOM) for power plants, ii) fixed operating and maintenance cost (FOM) for power plants, transmission lines, pipelines, and storage iii) nuclear fuel cost, and iv) gas and low-carbon fuel import cost.

\begin{figure}[htbp]

\begin{minipage}{.5\textwidth}
\centering
  \includegraphics[width=0.75\linewidth]{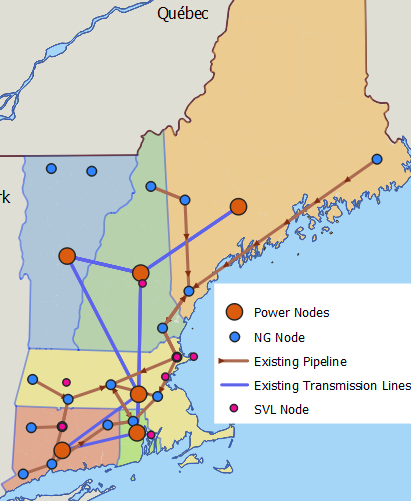}
  \caption*{(a)}
\end{minipage}%
\begin{minipage}{.5\textwidth}
\centering
  \includegraphics[width=0.9\linewidth]{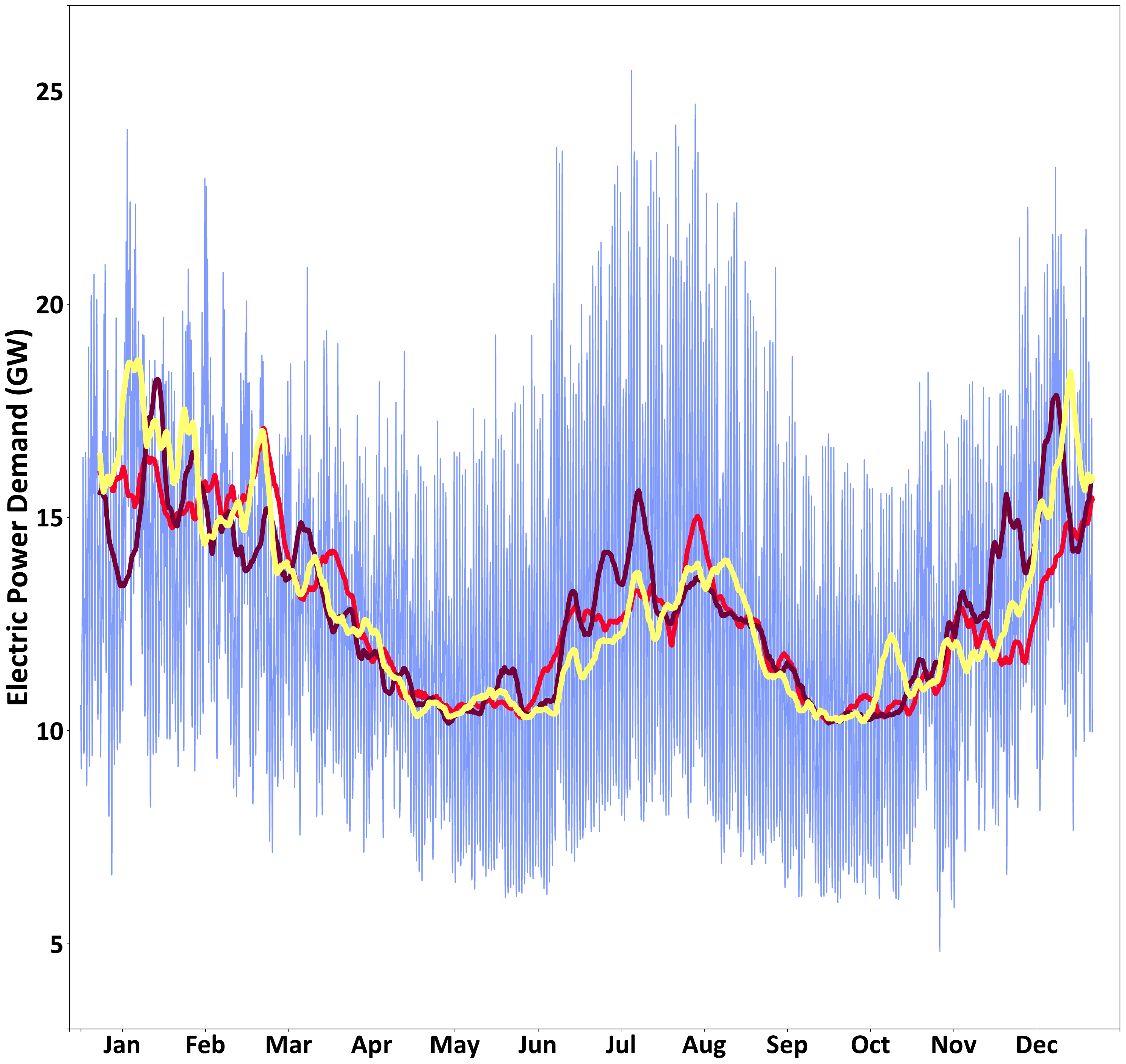} \caption*{(b)}
\end{minipage}
\caption{{Case study topology and demand variation.} (a): Existing topology for the joint power-gas system adopted from \citet{KhorramfarEtal2024_AE}; (b): Power demand variation case with three smoothed curves under high electrification (HE). Light blue lines show hourly power demand values for all 20 weather scenarios overlaid on top of each other to show the hourly variations across scenarios. Yellow, red, and brown bold lines illustrate rolling weekly averages of power demand for three of the 20 weather scenarios to highlight the impact of weather variation on scenario realizations.}
\label{fig:toplogy+load-variation}
\end{figure}

We consider the case study of New England to explore electricity and gas infrastructure investments and operations under future scenarios with stringent CO$_2$ emissions constraints and varying degrees of end-use electrification in the building sector. 
The New England region's spatial representation in the model, shown in Fig.~\ref{fig:toplogy+load-variation}a, consists of: a) a power network with 6 nodes (one node per state) connected by 23 exiting transmission lines and 40 candidate lines, b) a gas network with 23 nodes connected by 28 existing and 36 candidate pipelines.  The power system has 8 generation types, of which 3 are existing generation types and the rest are new ones. Existing power plant types include `ng' (gas-fired plants), `hydro',  and `nuclear'. The new plant types are `CCGT' (combined cycle gas turbine), `CCGT-CCS' (CCGT with carbon capture and storage technology that can capture 90\% of the resulting emissions),  `solar-new', `wind-new', and `wind-offshore'. The storage system is assumed to be the short-duration Lithium-ion grid-scale batteries. The gas infrastructure is served by supplying gas from two sources: a) NG with cost of 5.45\$/MMBtu depending on supply source and b) low-carbon fuel, representing synthetic or biomass-derived methane, that can be used with the existing infrastructure but comes at a substantial cost premium (20 \$/MMBtu vs. 5.45 \$/MMBtu for NG).  The low-carbon fuel is assumed to have zero CO$_2$ emissions owing to the biogenic nature of the CO$_2$ source. The full description of the model is given in Appendix~\ref{model-sp}.

The model outcomes are evaluated for 3 end-use electrification levels and two different decarbonization levels (detailed in \citet{Khorramfar2025_CRS}). 
The end-use electrification levels represent a different degree of electrification for the residential building sector in the New England region of the US, with the lowest electrification corresponding to the reference case (RF), followed by medium (ME) and high (HE) electrification levels. We also consider 80\% and 95\% decarbonization targets as all New England states have an emissions reduction goal of 80\% by 2050 below the baseline year of 1990 with some having a goal as high as 95\%. 

The reference \citep{Khorramfar2025_CRS} provides 20 projections of supply and demand data for the year 2050 for each of the electrification levels. The authors develop a \textit{bottom-up} model that outputs a projection based on historical supply and demand data, among others. Therefore, the number of projections is limited by the available historical data for the regions, in this case 20 years.  Note that each projection is treated as a realization of uncertainty. 
For each realization, the supply data consists of the capacity factors for solar, onshore wind, and offshore wind generators, and the demand data includes the bulk electric power and gas demand.  The realization of uncertain parameters show significant interannual variation as illustrated in the case of electricity demand in Fig.~\ref{fig:toplogy+load-variation}b. 

We set the value of lost load for power and gas systems at 10,000 \$/MWh and \$10,000 \$/MMBtu, respectively. The value of all other parameters is adopted from \citep{KhorramfarEtal2024_AE} and \citep{Khorramfar2025_CRS}.
Our implementation is based on JPoNG code in Python using the Gurobi solver.
All instances are run on the MIT Supercloud system with an Intel Xeon Platinum 8260 processor, up to 48 cores, and 192 GB of RAM \citep{Supercloud2018}. 

\section{Computational Experiments} \label{sec:comp-experiments}
The discussions of the numerical experiments are organized to explore distinct elements of the proposed model and solution approach, and compare the outcomes with SP, a common uncertainty characterization paradigm in energy infrastructure and operational planning. Appendix~\ref{App-sec:SP-model} provides the SP formulation and its model description.  
{In particular, computational experiments explore six aspects i) benchmarking the computational performance and solution quality of the SCM approach; ii) quantifying the importance of planning under uncertainty using well-known metrics in the literature; iii) comparing the co-optimize power-gas model with the power-only system; iv) isolating the impact of supply and demand uncertainty on DRO planning model outcomes, v) sensitivity of results to risk aversion of the energy planner, and vi) the out-of-sample performance of DRO and SP-based planning modeling frameworks. }

In the SP model, we assume equiprobable scenarios due to the lack of any further information on the frequency of scenarios. Throughout the experiments, the value of $\kappa$ in the MDRO model is et to 1. This value controls the weight of spatial correlation in the deviation from the nominal mean in the ambiguity set definition. {Our sensitivity analysis on $\kappa$ value in Appendix~\ref{app:kappa} suggests that the problem outcome is quite robust to $\kappa$ value and that the deviation terms in the ambiguity set definition have a limited role in impacting the problem outcome.} 
In the WDRO model, we set the value of Wasserstein distance, $\mathfrak{D}$ to the maximum element-wise distance between realizations of $\mathbb{P}$ and $\mathbb{Q}$, i.e., $\mathfrak{D}=\max \{||\omega_i-\zeta_j||, i\in \mathcal{M}, j\in \mathcal{K}\}$. Furthermore, unless otherwise stated, $\mathbb{P}$ contains 15 and $\mathbb{Q}$ contains 5 scenarios.

\subsection{Performance of the SCM}
This section evaluates the performance of SCM on the problem instances. Table~\ref{tab:SCMperformance} shows the performance of SCM against the extensive form (EF, i.e., solving MILP, problem~\eqref{model:DRO-EF}, with Gurobi directly) approach for the RF electrification level instances under different emissions targets,  number of representative days, and number of weather scenarios.  Although the strategic nature of the problem warrants patient planning, the solution time limit is set to 24 hours as a reasonable timeframe is required to conduct a range of sensitivity analysis.  
{For MDRO instances, the EF cannot achieve the optimal solution in 6 out of 16 instances within the time limit, whereas SCM achieves solutions with an average of 0.7\% gap within less than an hour. There are 3 out of 12 instances unsolved in WDRO's extensive form. Note that WDRO needs more than 5 scenarios to exceed the scenario number in the set $\mathcal{L}$. The average gap of the algorithms significantly increases in WDRO to 2.4\% but the solution time drops to around 15 minutes. The extensive form of SP instances poses much less computational burden, so all instances are solved with an average of about 2 hours. The performance of SCM in SP is comparable to MDRO with 0.76\% gap, but in a lower timeframe of 15 minutes. The performance of EF for higher numbers of representative days is not reported here as its performances greatly diminish and provides no baseline to calculate the gap.}

\begin{table}[htbp]
\small 
\centering
\caption{The performance of the SCM approximation method vs the extensive form (EF) solution. The 
CPU time is set to 24 hours for all instances. The gap is calculated by the difference between EF and SCM solutions, and normalized by the EF solution. {The WDRO instances start with 10 scenarios at we keep 5 scenarios in set $\mathcal{L}$ and the rest in $\mathcal{M}$.}}
\label{tab:SCMperformance}
\setlength{\tabcolsep}{3pt}
  \renewcommand{\arraystretch}{0.95} 
\begin{tabular}{c|c|c|ccc|ccc|ccc}
\toprule
\multicolumn{3}{c}{} & \multicolumn{3}{c}{MDRO}& \multicolumn{3}{c}{WDRO}&\multicolumn{3}{c}{SP}\\
\midrule 
\begin{tabular}[c]{@{}c@{}}Emis.\\ goal\end{tabular} & \begin{tabular}[c]{@{}c@{}}Rep. \\ days\end{tabular} & $|\mathcal{S}|$ & \begin{tabular}[c]{@{}c@{}}EF \\ CPU(h) \end{tabular} & \begin{tabular}[c]{@{}c@{}}SCM \\ CPU (h)\end{tabular} & \begin{tabular}[c]{@{}c@{}}Gap \\ (\%)\end{tabular}& \begin{tabular}[c]{@{}c@{}}EF \\ CPU(h)\end{tabular} & \begin{tabular}[c]{@{}c@{}}SCM \\ CPU (h)\end{tabular} & \begin{tabular}[c]{@{}c@{}}Gap \\ (\%)\end{tabular}&\begin{tabular}[c]{@{}c@{}}EF \\ CPU(h)\end{tabular} & \begin{tabular}[c]{@{}c@{}}SCM \\ CPU (h)\end{tabular} & \begin{tabular}[c]{@{}c@{}}Gap \\ (\%)\end{tabular}\\ 
\midrule 
\multirow{8}{*}{80\%} & \multirow{4}{*}{5} & 5 & 0.3& 0.03 & 0.93 && &&0.08&0.02&0.64 \\
& & 10& 2.1& 0.18 & 0.27 & 0.55 & 0.04  & 0.64&0.06&0.05& 0.3\\
& & 15& 13.31& 0.14 & 0.01& 4.15 & 0.14  & 1.18 &0.34&0.07&0.26\\
& & 20& 20.54& 1.47 & 1.54& 15.08 &0.16   &1.01  &1.05&0.48&0.85\\ \cmidrule{3-12}
& \multirow{4}{*}{10}& 5 & 0.51& 0.1 & 1.08 & & & &0.15&0.03&0.76 \\ 
& & 10& -& 0.33 &- & 4.07 & 0.18  & 5.55&1.89&0.08&1.15\\
& & 15& -& 1.48 & -& - & 0.21  & -&6.29&0.67&1.25 \\
& & 20& -& 1.25 & -& - & 0.81  & -&7.29&0.54&1.16 \\ \cmidrule{1-12}
\multirow{8}{*}{95\%} & \multirow{4}{*}{5} & 5 & 0.34& 0.03 & 1.05 & &&&0.09&0.02&1.34 \\
& & 10& 1.28& 0.22 & 0.17& 0.51 & 0.04  & 1.24&0.06&0.05&0.01 \\
& & 15& 8.86& 0.13 & 0.12& 3.14 & 0.11  & 1.38&1.16&0.06&0.18 \\
& & 20& 20.36 & 0.52 & 1.25 & 1.16 & 0.06  & 0.18 &1.67&0.39&0.42\\ \cmidrule{3-12}
& \multirow{4}{*}{10}& 5 &2.26& 0.12 & 0.63 & && &0.048&0.04&0.94 \\
& & 10& -& 1.21 & -& 3.76 & 0.16  & 4.62 &0.64&0.17&0.89 \\
& & 15& -& 1.03 & -& 7.64 & 0.2  & 4.63 &4.9&0.74&0.52\\
& & 20& -& 0.65 & -& -& 0.86  & -&2.94&0.56&1.55\\ 
\midrule 
Ave.  & && 6.98& 0.56 & 0.7& 6.02 & 0.26  & 2.4 &1.82&0.24& 0.76\\
\bottomrule 
\end{tabular}
\end{table}

\subsection{{Importance of Uncertainty}}
{To show the importance of considering uncertainty in weather-induced parameters, we calculate the Value of Stochastic Programming (VSS) and the Expected value of the Perfect Information (EVPI). Both of these measures are common criteria in the stochastic programming literature each revealing an aspect of using stochastic models over deterministic solutions \citep{Birge2011_SP_Book}.  VSS quantifies the expected gain from solving an SP compared to its deterministic counterpart. VSS is calculated in two steps. In the first step, the problem is solved with all random variables replaced by their statistical means. In the second step, the SP is solved such that the value of the first stage variables are set to their values in the first step \citep{ZhaoEtal2017_2SP_power_gas}.  In case of little to no uncertainty, the VSS approaches zero, indicating that the deterministic solution is a good approximation for the problem. }

{Future planning models often rely on prediction and projection methods to estimate the possible realization of random parameters. EVPI is a measure to quantify the value of perfect tools to determine the exact realization of scenarios. In other words, EVPI is the upper bound for the price that the planner should be willing to pay in the long run in order to have access to precise outcomes of the uncertain parameters \citep{Birge2011_SP_Book}. EVPI is calculated by solving the deterministic case for each scenario independently and then aggregating the objective values by their corresponding probabilities. }

{Table~\ref{tab:VSS_EVPI} shows the normalized values for VSS and EVPI. The solution of VSS is normalized by dividing the difference between the VSS solution and the original SP solution by the SP solution. The EVPI values are normalized in a similar fashion. With values ranging from 178 to 379\%, the VSS shows that the use of stochastic programming is highly recommended, especially under higher electrification levels. It is worth noting that the VSS values largely depend on the value of load shedding for the power system.  
This is because once the first stage variables are fixed, the system results in a solution with substantial load shedding, hence the higher the load shedding cost, the higher the VSS. As such, reducing or increasing the load-shedding cost for power and gas systems can significantly impact the values. However, the amount of load shedding indicates that the deterministic model can lead to a system that is severely unreliable for some weather outcomes.
The EVPI values indicate the share of the SP objective values that the planners should be willing to pay for perfect information. Since the total energy cost in SP solutions is very high (in our instances, around 20 billion dollars), these values are considerable, ranging from 0.5 to 1\$ billion. Overall, considering SP, and DRO models by extension, is highly recommended for energy systems planning. }

\begin{table}[htbp]
\centering 
\caption{Normalized VSS and EVPI. All values are in percentage. }
\label{tab:VSS_EVPI}
\setlength{\tabcolsep}{3pt}
  \renewcommand{\arraystretch}{0.95} 
\begin{tabular}{c|cccccc}
\toprule
Electrification level& \multicolumn{2}{c}{RF} & \multicolumn{2}{c}{ME} & \multicolumn{2}{c}{HE} \\
\midrule 
Decarbonization Target& 80\%& 95\% & 80\%& 95\% & 80\%& 95\% \\
\midrule 
VSS& 274&178&218& 325 & 274& 379 \\
EVPI  & 3.8& 2.7&5&4.8&3.8&4.4 \\
\bottomrule
\end{tabular}
\end{table}

\subsection{{Impact of Joint Energy System Optimization}}\label{sec:joint-vs-power}
{Energy transition affects the interdependency between power and gas systems. Here, the power system outcomes for the joint system versus the power-only model are compared to highlight the role of the joint planning model. The power-only model consists of variables and constraints pertaining to the electric power system. The emissions limit in this model is imposed on the carbon emissions resulting from power generation. Unlike the joint system in which the decarbonization goal is applied to the combined emissions of power and gas systems, the emissions reduction goal here is only applied to the power system, preventing carbon trading between two energy vectors. In addition to the coupling constraints, the joint system includes the added uncertainty in the gas load.} 

{Fig.~\ref{fig:pow-vs-powgas} shows the change of capacity, average generation, and cost components between power-gas system and power-only system for both MDRO and WDRO. The modeling paradigm and decarbonization target are the primary determinants of the difference between the two solutions. In the MDRO instances, offshore wind and CCGT-CSS generators replace the existing gas and storage under 80\% target, whereas the existing gas replaces the offshore wind and storage under 95\% target. Generation, however, is almost identical for all DRO instances between power-only and power-gas systems. The impact of increased renewable and costlier CCGT-CCS generators under 80\% target is reflected in the cost components such that the cost of investment and fixed cost (FOM) of generators and storage ('Power/storage inv+Fix') is higher for power-only system. In contrast, the cost of MDRO-95\% instances is lower for the power-only system instances due to less investment in renewables and storage. Overall, for the power-only system, the total power system cost is higher by 7.9\%, 6.1\% and 2.6\% under 80\% target, and lower by 9.2\%, 7.9\% and 14.6\% under 95\% target for RF, ME and HE electrification levels. }

{Similar to MDRO, the existing gas capacity reduces at 80\%  and increases at 95\% target for power-only WDRO instances. Unlike MDRO, the generation from less emission-intensive CCGT-CCS plants substantially replaces the offshore wind and conventional CCGT generation. As evident from Fig.\ref{fig:pow-vs-powgas}c, the investment and fixed cost increases at 80\% and decreases at 95\%. The impact of power-only approach on the load shedding is mixed. The load shedding increases at RF, and HE instances and decreases at ME.  Overall, for the power-only system, the total power system cost is higher by 7.7\%, 5.8\% and 2.3\% under 80\% target, and lower by 8\%, 8.4\% and 12.4\% under 95\% target for RF, ME and HE electrification levels. }

{
\begin{figure}
\centering
\includegraphics[width=1.1\textwidth]{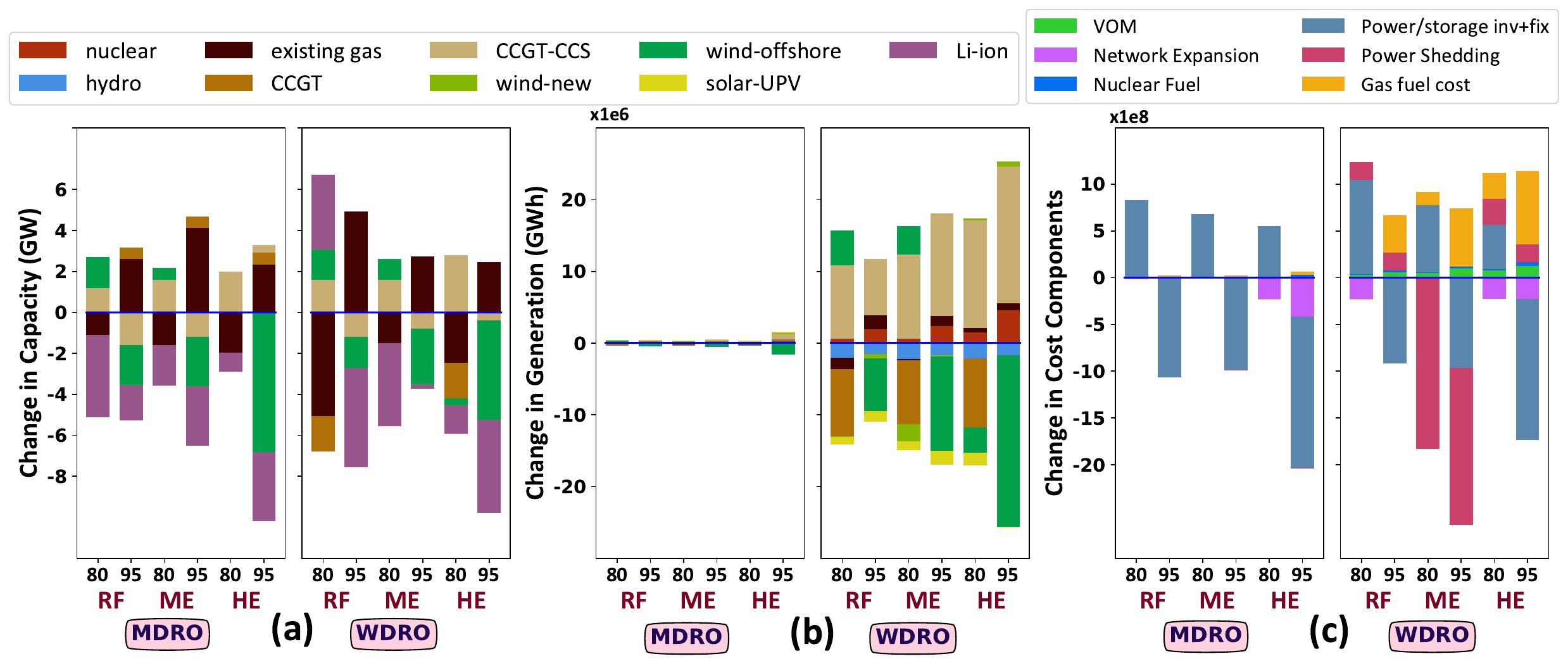}
\caption{Difference in capacity (left) and cost components (right) between joint power-gas model vs. power-only model}
\label{fig:pow-vs-powgas}
\end{figure}}

\subsection{Impact of Demand and Supply Uncertainty}


Often time the literature overlooks the uncertainty in all the key parameters affected by weather variations (see Table~\ref{tab:lit-review}). 
We carry out experiments to highlight the importance of uncertainty in both supply and demand parameters and their impact on key energy infrastructure planning outcomes of interest. Specifically, the DRO and SP models are evaluated for three cases where uncertainty is modeled as follows: a) uncertainty only in supply parameters, i.e. VRE capacity factors (S1D0), b) uncertainty in electricity and gas demands only (S0D1), and c) uncertainty in both supply and demand parameters (S1D1).

\begin{figure}[htbp]
\centering
\includegraphics[width=\textwidth]{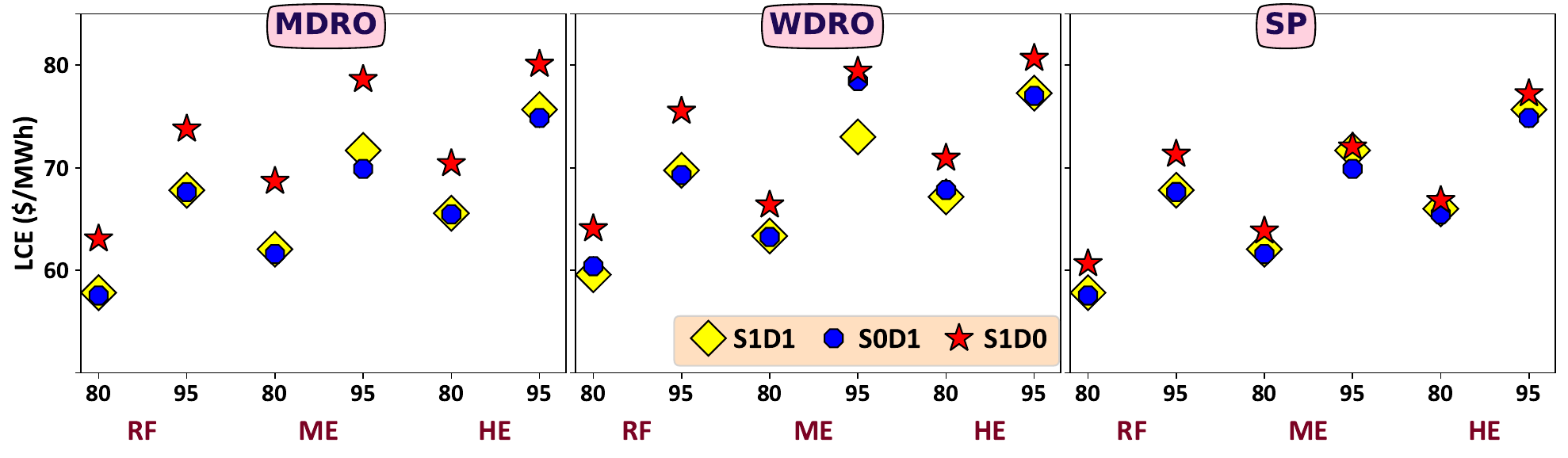}
\caption{Levelized Cost of Energy (LCE) for instances where both supply and demand are uncertain (S1D1), only supply is uncertain (S1D0), and only demand is uncertain (S0D1). LCE is calculated by dividing the total system cost by the average combined power-gas demand across scenarios.}
\label{fig:levelized-cost}
\end{figure}

{The complementarity effect between various generators and across different potential locations is a subject of some studies \citep{BergerEtal2022_OptimLetter, QiuEtal2024_CRS}. Weather variation affects both supply and demand parameters. Therefore, ignoring the uncertainty in one set of parameters can distort the complementarity between weather-sensitive parameters. This section isolates and evaluates the impact of uncertainty in either supply or demand parameters.   
Fig.~\ref{fig:levelized-cost} shows the levelized cost of energy (LCE) under various parameter uncertainty options. Here, we define LCE as the ratio of the total cost to average combined power-gas demand \footnote{For each scenario, gas demand is converted to MWh unit and summed with power demand, before taking the average across scenario realizations}. In S0D1 and S1D0 instances, the deterministic supply or demand parameter is taken from the scenario with the closest aggregate power-gas demand to the median of all scenarios.  
Under supply parameters uncertainty (S1D0), all instances tend to overestimate the total system cost, 5.8-10.7\% in MDRO, 4.4-8.8\% in WDRO, and 0.5-5.2\% in SP as compared to S1D1 instances. Interestingly, the cost overestimation slightly decreases in the higher emissions limit. 
In contrast, under demand-only uncertainty (S0D1), the cost deviation is not substantial from S1D1 for all three modeling paradigms. There is a slight underestimation in MDRO and SP by as low as 3\%; the deviation for WDRO is mixed between -1 and 7\%. Therefore, in case of using MDRO and SP, demand uncertainty is a better proxy than supply uncertainty to account for weather-induced uncertainty. }


\subsection{Risk Analysis}
Various entities can use energy models for better-informed decisions. These entities can include utility companies, balancing authorities, policymakers and government agencies. Each entity, however, may manifest a differing sensitivity toward the worst-case realization, leading to different levels of risk-aversion. Table~\ref{tab:risk} presents the difference between risk-neutral and risk-averse instances. {The instances cover all three electrification levels and uncertainty modeling paradigms under a mild ($\lambda=0.5, \alpha=0.7$) and strong ($\lambda=0, \alpha=0.9$) risk-aversion. 
In all instances, the risk aversion of the decision maker leads to a higher total cost of up to 9.5\% for MDRO, 295\% for WDRO and 9.5\% for SP. The risk-sensitivity of WDRO is significantly higher such that a mild risk-aversion can increase the total cost by 22\% on averages. Interestingly, ME presents the highest sensitivity across electrification levels, indicating that high electrification is not necessarily the most ris-sensitive pathway.}

Table~\ref{tab:risk} also shows the breakdown of the solutions per major system outcomes. {While the risk measure consistently results in a higher total system cost, its impact on key problem outcomes can be mixed. In MDRO, the storage capacity decreases in mild risk aversion and increases otherwise. Thermal generation increases in all electrification replacing solar, and in some instances, wind generation. Similar observations can be made for the SP outcomes. The pattern for WDRO is more consistent such that the increased wind generation (both from onshore and offshore) replaces the thermal and solar generation. Overall, this interplay leads to increased VRE capacity, and increased storage capacity in all but one instance (RF, $\lambda=0.5$, $\alpha=0.5$). }


\begin{table}[htbp]
\small
\centering 
\caption{The percentage change in key problem outcomes between risk-averse and risk-neutral instances ($\lambda=1$) under different electrification levels. The emissions reduction goal of 80\% is applied to all instances. }
\label{tab:risk}
\setlength{\tabcolsep}{4pt}
  \renewcommand{\arraystretch}{0.9} 
  {
\begin{tabular}{cccc|lllllll}
\toprule 
\multicolumn{1}{l}{\begin{tabular}[c]{@{}l@{}}Elec.\\Level\end{tabular}} & \multicolumn{1}{l}{Model} & $\lambda$ & $\alpha$ & \begin{tabular}[c]{@{}l@{}}Total \\cost\end{tabular} & \begin{tabular}[c]{@{}l@{}}Storage\\cap.\end{tabular} & \begin{tabular}[c]{@{}l@{}}Thermal\\cap.$^\ast$\end{tabular} & \begin{tabular}[c]{@{}l@{}}VRE\\cap\end{tabular} & \begin{tabular}[c]{@{}l@{}}Thermal\\gen.$^\ast$\end{tabular} & \begin{tabular}[c]{@{}l@{}}Solar\\gen.\end{tabular} & \begin{tabular}[c]{@{}l@{}}Wind\\gen.$^\ast$\end{tabular} \\
\midrule
\multirow{6}{*}{RF}& \multirow{2}{*}{MDRO}& 0.5& 0.5&0.5& -5.3&1.5& 0.3& 0.9& -1.3& 0.1 \\
 && 0  & 0.9 & 2.3&	2.3&	-6.6&	-0.2&	7.4	&-5.1&	-3.4 \\
  \arrayrulecolor{gray}\cmidrule{5-11}
 & \multirow{2}{*}{WDRO}& 0.5& 0.5& 22&	-8&	3&	11&	-20&	-9&	28\\
 && 0  & 0.9 &295&	72&	-36&	72&	-61&	-23&	87\\
 \arrayrulecolor{gray}\cmidrule{5-11}
 & \multirow{2}{*}{SP}& 0.5& 0.5& 0.8&	-4&	0.5&	0.1&	-0.3&	-0.1&	0.4  \\
 && 0  & 0.9 &2.3&	2.3&	-6.6&	-0.2&	8.2&	-9.5&	-2.6 \\
 \midrule
\multirow{4}{*}{ME}& \multirow{2}{*}{MDRO}& 0.5& 0.5& 2.2&	5.6&	-2&	1.1&	0.4&	-2.9&	1.6\\
 && 0  & 0.9 &9.5&	-9.6&	-6.4&	0.5&	6.8&	-4.8&	-3.4 \\
 \arrayrulecolor{gray}\cmidrule{5-11}
 & \multirow{2}{*}{WDRO}& 0.5& 0.5& 18&	26&	-9&	8&	-16&	-4&	18\\
 && 0  & 0.9 &236&	81&	-34&	64&	-60&	-9&	75 \\
 \arrayrulecolor{gray}\cmidrule{5-11}
 & \multirow{2}{*}{SP}& 0.5& 0.5&1.8&	2.8&	-1&	1&	-2.5&	-0.8&	2.6 \\
 && 0  & 0.9 &9.5&	-9.6&	-6.4&	0.5&	4.7&	-3.3&	-1.8 \\
 \midrule
\multirow{4}{*}{HE}& \multirow{2}{*}{MDRO}& 0.5& 0.5& 1&	-4.1&	1.3&	0&	2&	0.3&	-1.7\\
 && 0  & 0.9 & 3.7&	-10&	3.3&	0.4&	8.2&	-8.7&	-2.3\\
 \arrayrulecolor{gray}\cmidrule{5-11}
 & \multirow{2}{*}{WDRO}& 0.5& 0.5&16&	37&	-5&	13&	-21&	-8&	25\\
 && 0  & 0.9 &163&	179&	-27&	68&	-63&	5&	63\\
 \arrayrulecolor{gray}\cmidrule{5-11}
 & \multirow{2}{*}{SP}& 0.5& 0.5& 0.7&	-1.3&	0.4&	0.1&	-0.2&	-0.2&	0.4\\
 && 0  & 0.9 &3.7&	-10&	3.3&	0.4&	9.2&	-11.4&	-2.3\\
 \bottomrule
\end{tabular}}
{\\ \small {$\ast$  `Thermal cap.' and `Thermal gen.' refer to the combined capacity and generation from nuclear and gas-fired power plants, respectively. Similarly, `Wind gen.' refers to the combined generation from both onshore and offshore wind plants. }}
\end{table}

\normalsize

\subsection{Out-of-sample Performance}
Our analyses so far relied on available supply-demand projections for the future target year. However, the projection methods as well as climate change trends can result in different parameters and possibly more extreme realizations. Therefore, in this section, we evaluate the out-of-sample performance of DRO and SP models. Our case study has 20 projections for supply and demand parameters that are randomly partitioned into 10 in-sample and 10 out-of-sample sets. First, the MDRO, WDRO, and SP models are applied to the in-sample scenarios to obtain investment decisions. Then the out-of-sample operational performance is evaluated using in-sample investment decisions. For the WDRO instances, we keep 2 of the scenarios in set $\mathcal{L}$ and the rest in $\mathcal{M}$.

\begin{figure}[htbp]
\centering
\includegraphics[width=0.7\textwidth]{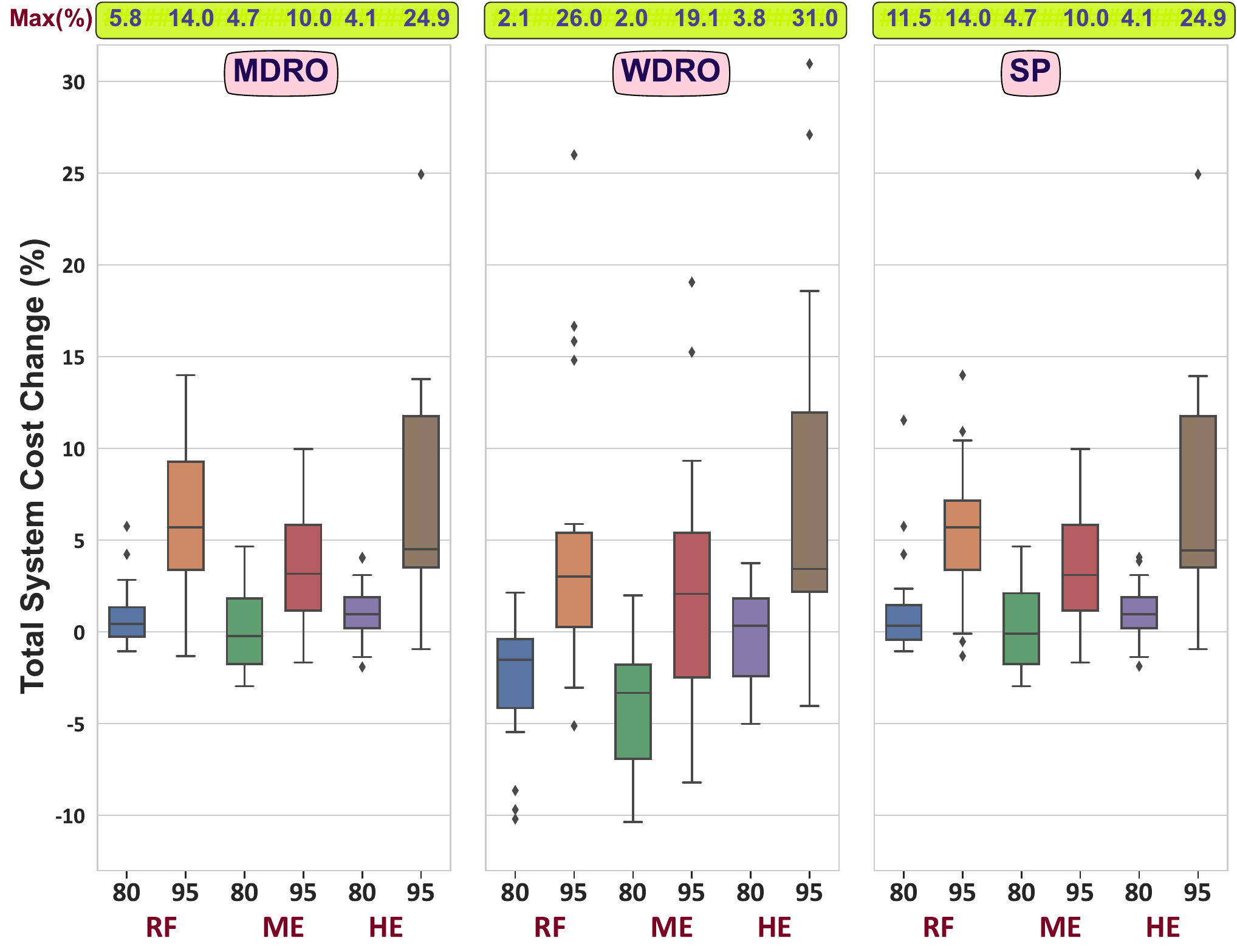}
\caption{Out-of-sample performance over 20 randomly generated partitions of the scenario space. Each partition is a random draw of 10 scenarios from the set of 20 scenarios. The selected scenarios are treated as in-sample, and the rest as out-of-sample scenarios. For each scenario, the system cost change in comparison to its corresponding in-sample cost is reported. The maximum increase of total cost between out-of-sample scenarios and their corresponding in-sample set is reported at the top of each plot. }
\label{fig:out-of-sample}
\end{figure}

The total system cost for out-of-sample scenarios over 20 random partitions of the scenario space, as well as the largest cost difference between in-sample and out-of-sample scenario sets, is depicted in Fig.~\ref{fig:out-of-sample}. 
{For all electrification levels and decarbonization targets, MDRO presents similar performance as SP, with the exception of RF-80\% in which MDRO has better performance. WDRO's performance is mixed. While it outperforms other models in all instances with 80\% target, its out-of-sample performance deteriorates under 95\% target compared to MDRO and SP.}
{The out-of-sample performance is considerably better for the 80\% decarbonization target for all models, as evident from the small box plot variance and maximum performance. This can be attributed to lower share of dispatchable power generation or insufficient storage capacity in more ambitious decarbonization targets.} 
{The maximum deterioration of total cost in 80\% target is 4.1-5.8\% in MDRO, 2-3.8\% in WDRO, and 4.1-11.5\% in SP suggesting that DRO modeling approaches can better hedge against uncertainty in less ambitious emissions limits. In contrast, the worst deterioration for 95\% target increase to 10-24.9\% for MDRO, 19.1-31\% for WDRO, and 10-24.9\% for SP, indicating the outperformance of MDRO and SP over WDRO.}  

\section{Conclusion }\label{sec:conclusion}
In this paper, we considered a joint planning of electric power and gas systems as a GTEP, and proposed a framework to characterize power and gas demand as well as the supply of renewable generation via distributionally robust optimization. Our modeling approach incorporated the risk sensitivity of the energy planner as conditional value-at-risk to hedge against the worst realization of parameters. We developed DRO models with both moment and metric-based (Wasserstein distance) ambiguity sets. 

In the moment-based model, we enriched the ambiguity set by the correlation between power-gas nodes, and showed that both models can be formulated as MILP problems. The resulting MILP is a large-scale model that is intractable to solve in larger sizes. Therefore, we developed an efficient approximation scheme that relies on solving a sequence of linear programs to construct a feasible solution.
Our modeling approach is applied to future energy planning of the New England region under different emissions reduction goals as well as levels of residential heating electrification which impacts the long-term projection of power-gas demands. Finally, our extensive numerical analyses highlight the challenges surrounding energy transition as pertained to weather-induced uncertainty and shed light on various aspects of the modeling approach.

{First, our analyses show that planning under uncertainty is indispensable to achieving a more reliable future energy system, otherwise the system may face insufficient generation capacity or misplaced assets leading to severe blackout events. Second, disregarding power system's interdependence with gas can distort outcomes, resulting in lower system cost in less ambitious decarbonization targets and vice versa for more stringent emissions limit. Third, the complementarity effect between capacity factors and demand is substantial, especially under DRO models. Although excluding either demand or supply uncertainty can lead to different outcomes, the results suggest that demand uncertainty is a better proxy for uncertainties posed by the weather.}  Fourth, the energy transition involves a vast array of stakeholders whose risk sensitivities should be considered as part of the modeling framework. Finally, the experiments highlighted the inherent uncertainty in weather-induced parameters and quantified the out-of-sample performance of each modeling approach. 

{From the modeling perspective, the theme of our study is two-fold. First, we stress the role of the modeling approach in characterizing uncertainty. While there are many paradigms to endogenize uncertainty, energy transition planners should opt for modeling tools that better describe the type of uncertainty and available data. Second, the scalability of energy planning models is a persistent challenge despite advances in recent years. Although the exact solution methods provide optimality guarantees, their timeframe can be prohibitively long to conduct a range of sensitivity analyses which is an important aspect in energy planning models. }

{Future directions are envisioned to investigate other key dimensions of the planning under uncertainty.  The DRO models developed for discrete realizations can be extended to handle continuous ambiguity sets that cover an ensemble of projections. The resulting model can present nonconvexity, therefore, future works can consider ML-assisted optimization models to not only streamline the solution process but also to generate synthetic projections with varying degrees of extremities. }
The approximate solution method presented here performs efficiently for a limited number of representative days, but it may fail in larger sizes. {A decomposition method that can utilize the problem structure can be developed to further address the problem's scalability.}

{Beyond the aforementioned directions, future endeavors can augment the modeling fidelity. This study adopts a central planner perspective, whereas in practice there are multiple entities such as government agencies, utility companies and balancing authorities that are involved in energy infrastructure planning. Therefore, a decentralized approach that captures the main interdependence between these entities can offer a higher fidelity model. Eventually, future studies can encompass uncertainties in other parameters as future energy planning is subject to myriads of unpredictabilities stemming from fluctuating prices, climate change trends, investment costs, regulations, and geopolitical factors among others.}



\section*{Acknowledgment}
This work was supported by MIT Energy Initiative Future Energy System Center, Department of Energy's Project EARNEST (An Equitable, Affordable \& Resilient Nationwide Energy System Transition), and MIT Climate Grand Challenge ``Preparing for a new world of weather and climate extremes''. The responsibility for the contents lies with the authors.

\appendix  

\section{Illustrative Example for the Moment-based Ambiguity Set} \label{app:example}

As an illustrative example, consider a network with 6 nodes. For node $i=1$, assume that the average demand correlation across time periods and scenarios and  inverse distance between node $i=1$ and all other 5 nodes are given as:
\begin{align*}
&[\rho^{\text{e}}_{i2}, \rho^{\text{e}}_{i3},\ldots, \rho^{\text{e}}_{i6}] = [0.9, 0.5, -0.1,0.8, -0.3]\\
& [\ell_{i2}, \ell_{i3}, \ldots, \ell_{i6}]= [\frac{1}{10}, \frac{1}{2}, \frac{1}{5}, \frac{1}{2}, \frac{1}{1}]
\end{align*}
{
For $\kappa=1$, the term $\kappa \ell_{ij}\rho_{ij}$ which is used to determine the deviations is calculated for nodes 2 to 6 as:
$$
\kappa \ell_{ij}\rho_{ij} = \{0.09, \ 0.25, \ -0.05,\  0.4,\  -0.3\}, \qquad  i=1, j\in \{2,3,\ldots,6\}
$$
Node $i=1$ has negative demand correlations with nodes 4 and 6 and positive correlations with the rest. Therefore, 
\begin{align*}
& \barbelow{\Delta}^{\text{e}}_i = \min\{-0.05,-0.3 \}=-0.3\\
& \bar{\Delta}^{\text{e}}_i = \max\{0.09, 0.25, 0.4\}=0.4 \\
\end{align*}}

\section{Proofs of Propositions}\label{app:proofs}
This section contains the proofs for all the propositions in the main text.

\subsection{Proof of Proposition \ref{prop1}}
\noindent \textbf{Proof}: We follow the proof presented in \citep{BasciftciEtal2023COR}. For any $\xeg\in \mathcal{X}$, let $(\boldsymbol{\phi}_1,\boldsymbol{\phi}_2, \boldsymbol{\phi}_3, \boldsymbol{\phi}_4)\in \mathbb{R}$ be the dual variables associated with constraints \eqref{val_func:c1} to \eqref{val_func:c4}, respectively.  Consider the dual of \eqref{model:val_func}:

\begin{align}\label{model:dual_val_function}
\max \ & \boldsymbol{\phi}_1(\de_s -\AeO \xe)+\boldsymbol{\phi}_2(\bT-\Fv_s\AeT \xe) \\
& + \boldsymbol{\phi}_3(\dg_s - \AgO\xg)+\boldsymbol{\phi}_4(\bb-\Ae\xe -\Ag\xg)\notag\\
\text{s.t.} & (\BeO)^T \boldsymbol{\phi}_1 + (\BeT)^T \boldsymbol{\phi}_2 + (\Be)^T \boldsymbol{\phi}_4 \leq \CeT \notag\\
& (\BgO)^T \boldsymbol{\phi}_3 + (\Bg)^T \boldsymbol{\phi}_4 \leq \CgT \notag
\end{align}

The feasible region of the dual problem \eqref{model:dual_val_function} is independent of the first stage variable $\xeg$ and the realization of the uncertain parameters. On one hand, its solution is always bounded as for every $ \xeg\in \mathcal{X}$ the problem has complete recourse. On the other hand, for a fixed dual variables $(\boldsymbol{\phi}_1,\boldsymbol{\phi}_2, \boldsymbol{\phi}_3, \boldsymbol{\phi}_4)$, the objective function of model \eqref{model:dual_val_function} is a linear function of $\xeg$ and scenario realization $\boldsymbol{\omega}_s$, hence is convex and piecewise linear. \halmos


\subsection{Proof of Proposition \ref{prop2}}

\noindent \textbf{Proof:} Consider $\max_{\mathbb{P}\in \mathcal{P}(\mathcal{S}, \kappa)}\mathbb{E}_{\mathbb{P}}[V(\xeg,\boldsymbol{\omega}_s)]$ term under the ambiguity set \eqref{ambguity_set} with dual variables indicated within brackets:

\begin{align} \label{model:inner_max_expect}
\max_{p} \ &\sum_{s\in S}p_s V(\xeg,\boldsymbol{\omega}_s)\\
\text{s.t.} \ & \sum_{s\in \mathcal{S}}p_s d^{\text{e}}_{sit} \leq \mu^{\text{e}}_{it}+\bar{\Delta}^e_{i}&  i \in \Ne, t\in \Te, \quad [\beta^1_{it}] \notag\\
& -\sum_{s\in \mathcal{S}}p_s d^{\text{e}}_{sit} \leq  -\mu^{\text{e}}_{it}-\underline{\Delta}^e_{i} &  i \in \Ne, t\in \Te, \quad [\beta^2_{it}]\notag\\
&\sum_{s\in \mathcal{S}}p_s d^v_{sit} \leq \mu^{v}_{it}+\bar{\Delta}^v_{i} & v\in \mathcal{V},  i \in \Ne, t\in \Te, \quad [\gamma^1_{vit}]\notag\\
&-\sum_{s\in \mathcal{S}}p_s d^v_{sit} \leq  -\mu^{v}_{it}-\barbelow{\Delta}^v_{i} & v\in \mathcal{V},  i \in \Ne, t\in \Te, \quad [\gamma^2_{vit}]\notag\\
&\sum_{s\in \mathcal{S}}p_s \dg_{sk\tau}\leq \mu^{\text{g}}_{k\tau}+\bar{\Delta}^g_{k} & k \in \Ng, \tau \in \Tg, \quad [\pi^1_{k\tau}]\notag\\
& -\sum_{s\in \mathcal{S}}p_s \dg_{s k\tau} \leq -\mu^{\text{g}}_{k\tau}-\barbelow{\Delta}^g_{k} & k \in \Ng, \tau \in \Tg, \quad [\pi^2_{k\tau}]\notag\\
& \sum_{s\in S} p_s =1 & i\in \mathcal{N}, \quad  [\xi]\notag\\
&p_s\in \mathbb{R}^{|S|} & s\in \mathcal{S}, i\in \notag \mathcal{N}
\end{align}

We write the dual of \eqref{model:inner_max_expect} as the following minimization problem:
\begin{subequations}\label{model:inner_prob_companct}
    \begin{align}
\min \quad &\Psi^{\text{obj}}  + \xi\\
\text{s.t. }\ & \xi \geq V(\xeg,\boldsymbol{\omega}_s)-\Psi^{\text{cntr}}_{s}, \qquad s\in \mathcal{S}\label{c1:inner_prob_compact}
\end{align}
\end{subequations}

where $\Psi^{\text{cntr}}_{s}$ and $\Psi^{\text{obj}}$ are defined as
\begin{align*}
\Psi^{\text{cntr}}_{s} =&  \sum_{i\in \mathcal{N}^{\text{e}}}\sum_{ t \in \mathcal{T}^{\text{e}}} \left [\beta^1_{it}d^{\text{e}}_{sit} -\beta^2_{it}d^{\text{e}}_{sit} \right] +\sum_{i\in \mathcal{N}^{\text{e}}}\sum_{ t \in \mathcal{T}^{\text{e}}}\sum_{v\in \mathcal{V}} \left [\gamma^1_{vit}d^{\text{v}}_{sit} -\gamma^2_{vit}d^{\text{v}}_{sit} \right] + \sum_{k\in \mathcal{N}^{\text{g}}} \sum_{ \tau \in \mathcal{T}^{\text{g}}} \left [\pi^1_{k\tau} d^{\text{g}}_{sk\tau} -\pi^2_{vk\tau} d^{\text{g}}_{sk\tau} \right], s\in S \notag \\
\Psi^{\text{obj}}= & \sum_{i\in \mathcal{N}^{\text{e}}, t \in \mathcal{T}^{\text{e}}}\left[ (\mu^{\text{e}}_{it}+\bar{\Delta}^e_i)\beta^1_{it}-(\mu^{\text{e}}_{it}+\barbelow{\Delta}^e_i)\beta^2_{it} 
+\sum_{v\in \mathcal{V}}( (\mu^{v}_{it}+\bar{\Delta}^v_i)\gamma^1_{vit}-(\mu^{v}_{it}+\barbelow{\Delta}^v_i)\gamma^2_{vit}) \right] \\
&+\sum_{k\in \mathcal{N}^{\text{g}}, \tau \in \mathcal{T}^{\text{g}}}\left[ (\mu^{\text{g}}_{k\tau}+\bar{\Delta}^{\text{g}}_{k})\pi^1_{k\tau}-(\mu^{\text{g}}_{k\tau}+\barbelow{\Delta}^{\text{g}}_{k})\pi^2_{k\tau} \right]
\end{align*}

The constrain of \eqref{c1:inner_prob_compact} has to be satisfied for all, including the largest scenarios (i.e., maximum). Therefore, problem~\eqref{model:inner_prob_companct} is equivalent to:
\begin{align}\label{model:expet_DRO}
\min \ & \Psi^{\text{obj}}+\max_{s\in \mathcal{S}} \{V(\xeg,\boldsymbol{\omega}_s)-\Psi^{\text{cntr}}_s\}
\end{align}

Now consider the second inner maximization term with \cvar{} measure:

\begin{align*}
\max_{\mathbb{P}\in \mathcal{P}(\mathcal{S}, \kappa)}\ \cvar^\alpha_{\mathbb{P}}(V(\xeg,\boldsymbol{\omega}_s))=\max_{\mathbb{P}\in \mathcal{P}(\mathcal{S}, \kappa)}\ \min_{\eta\in \mathbb{R}} \{\eta +\frac{1}{1-\alpha}\mathbb{E}_{\mathbb{P}}[\max(V(\xeg,\boldsymbol{\omega}_s)-\eta,0)] \} 
\end{align*}
Note that the term inside the minimization expression is convex on $\eta$ \citep{SarykalinEtal2008_CVaR_VaR}, and the  \cvar{} is concave on $\mathbb{P}$ \citep{RahimianMehrotra2019_DRO_review}. Therefore, the order of maximum and minimum operators can exchange \citep{Sion1958_MinMax} to yield:

\begin{align}\label{model:cvar_DRO}
\max_{\mathbb{P}\in \mathcal{P}(\mathcal{S}, \kappa)}\ \cvar^\alpha_{\mathbb{P}}(V(\xeg,\boldsymbol{\omega}_s))=\min_{\eta\in \mathbb{R}} \{\eta +\frac{1}{1-\alpha}\max_{\mathbb{P}\in \mathcal{P}(\mathcal{S}, \kappa)} \mathbb{E}_{\mathbb{P}}[\max(V(\xeg,\boldsymbol{\omega}_s)-\eta,0)] \} 
\end{align}
Following a similar approach for first maximization term, we can write \eqref{model:cvar_DRO} as:

\begin{subequations}
\begin{align*}
\min \quad & \eta +\frac{1}{1-\alpha}(\Psi^{\text{obj}}+\xi)\\
\text{s.t. }\ & \Psi^{\text{cntr}}_s+\xi \geq V(\xeg,\boldsymbol{\omega}_s)-\eta, \qquad  s\in S\\
& \Psi^{\text{cntr}}_s+\xi \geq 0, s\in S
\end{align*}
\end{subequations}
Since the first set of constraints has to be satisfied for every $s \in \mathcal{S}$, we equivalently have:
\begin{subequations}
\begin{align*}
\min \quad & \eta +\frac{1}{1-\alpha}(\Psi^{\text{obj}}+\xi)\\
\text{s.t. }\ & \eta \geq \max_{s\in S} \{V(\xeg,\boldsymbol{\omega}_s)- \Psi^{\text{cntr}}_s-\xi\}\\
& \Psi^{\text{cntr}}_s+\xi \geq 0, s\in S
\end{align*}
\end{subequations}

Minimizing the right-hand-side of the first constraints is equivalent to minimizing $\eta$, therefore:

\begin{align*}
\min \quad & \max_{s\in S} \{V(\xeg,\boldsymbol{\omega}_s)- \Psi^{\text{cntr}}_s\} + \frac{\alpha \xi}{1-\alpha} +\frac{\Psi^{\text{obj}}}{1-\alpha}\\
\text{s.t. }\ &\Psi^{\text{cntr}}_s+\xi \geq 0, s\in S
\end{align*}

Then, the model \eqref{model:dro_orig} becomes:

\begin{align*} 
\min_{x\in \mathcal{X}} \ & \CeO \xe + \CgO \xg \\
& + \lambda \left[\Psi^{\text{obj}} +\max_{s\in S} \{V(\xeg,\boldsymbol{\omega}_s)- \Psi^{\text{cntr}}_s\}\right] \\
& + (1-\lambda)\left[(\max_{s\in S} \{V(\xeg,\boldsymbol{\omega}_s)- \Psi^{\text{cntr}}_s\}+\frac{\alpha \xi}{1-\alpha}+\frac{\Psi^{\text{obj}}}{1-\alpha}\right]\\
\text{s.t. }\ &\Psi^{\text{cntr}}_s+\xi \geq 0& s\in S
\end{align*}
\noindent which is equivalent to the following MILP:
  \begin{subequations}\label{model:DRO-EF2}
\begin{align}
\min \ & \CeO \xe + \CgO \xg + \lambda \Psi^{\text{obj}} +\delta  + (1-\lambda)\left[\frac{\alpha \xi}{1-\alpha}+\frac{\Psi^{\text{obj}}}{1-\alpha}\right]\\
\text{s.t. }\ &\Psi^{\text{cntr}}_s+\xi \geq 0& s\in S\\
& \delta \geq \CeT \ye_s+ \CgT\yg_s- \Psi^{\text{cntr}}_s & s\in S\\
&\eqref{val_func:c1}-\eqref{val_func:c4}\\
& \xe,\xg \in \mathbb{Z}^+ \times  \mathbb{R}^+,\ye_s,\yg_s, \xi \in \mathbb{R}\\
& \boldsymbol{\beta}^1, \boldsymbol{\beta}^2,\boldsymbol{\gamma}^1, \boldsymbol{\gamma}^2, \boldsymbol{\pi}^1,\boldsymbol{\pi}^2 \in \mathbb{R}^+
\end{align}
\end{subequations}
\hfill  \halmos

\subsection{Proof of Proposition \ref{prop3}}

Based on the ambiguity set defined in model \eqref{model:wass}, the inner maximization term becomes
\begin{align}
    \max_{\mathbb{P}\in \mathcal{P}(\mathbb{Q}, \mathfrak{D})} &\sum_{i\in \mathcal{M}}p_i V(x,\omega)\\
    \text{s.t.}\ & W_L(\mathbb{P},\mathbb{Q}) \leq \mathfrak{D}
\end{align}

\noindent or equivalently

\begin{subequations}
    \begin{align}
    \max_{\mathbb{P}\in \mathcal{P}(\mathcal{S})} &\sum_{i\in \mathcal{M}}p_i V(x,\omega_i)\\
    \text{s.t.}\ & \sum_{j\in \mathcal{K}}\sum_{i\in \mathcal{M}} ||\omega_i-\zeta_j||^L \pi_{ij} \leq \mathfrak{D} & \label{wass3-c1}\\
    & \sum_{j\in \mathcal{K}} \pi_{ij}-p_i = 0 & i\in \mathcal{M} \label{wass3-c2} \\\
    & \sum_{i\in \mathcal{M}} \pi_{ij} = q_j & j \in \mathcal{K}\ \label{wass3-c3}  \\
    & \sum_{i\in \mathcal{M}} p_i =1& \label{wass3-c4} \\
    &\pi_{ij}, p_i\in \mathbb{R}^+
\end{align}
\end{subequations}

Based on constraints \eqref{wass3-c2}, $p_i$ can be eliminated to achieve
\begin{subequations}\label{model:wass-model1}
    \begin{align}
    \max_{\mathbb{P}\in \mathcal{P}(\mathcal{S})} &\sum_{i\in \mathcal{M}}\sum_{j\in \mathcal{K}} \pi_{ij} V(x,\omega_i)\\
    \text{s.t.}\ & \sum_{j\in \mathcal{K}}\sum_{i\in \mathcal{M}} ||\omega_i-\zeta_j||^L \pi_{ij} \leq \mathfrak{D} & \ [\beta^1]\\
    & \sum_{i\in \mathcal{M}} \pi_{ij} = q_j & j \in \mathcal{K}\ [\beta^2_j]\\
    &\pi_{ij}\in \mathbb{R}^+
\end{align}
\end{subequations}

{Note that constraint~\eqref{wass3-c4} is no longer needed because $\sum_{i,j}\pi_{ij}$ becomes equal to $\sum_{j}q_j$ according to constraints~\eqref{wass3-c3} which is equal to 1 by definition.}
The dual of model \eqref{model:wass-model1} is
\begin{subequations}\label{model:wass-expect}
    \begin{align}
        \min \ &\ \mathfrak{D}\beta^1+\sum_{j\in \mathcal{K}} \beta^2_j q_j\\
     \text{s.t.}\  & ||\omega_i-\zeta_j||^L \beta^1 + \beta^2_j \geq V(x,\omega_i) & i\in \mathcal{M}, j\in \mathcal{K} \label{wass-c1}\\
       & \beta^1\in \mathbb{R}^+, \beta^2_i \in \mathbb{R} \label{wass-c2}
    \end{align}
\end{subequations}

Applying the same procedure, the second minimization term (i.e., risk term) becomes

\begin{subequations}\label{model:wass-cvar}
    \begin{align}
        \min \ &\ \eta+\frac{1}{1-\alpha}(\mathfrak{D}\beta^1+\sum_{j\in \mathcal{K}} \beta^2_j q_j)\\
     \text{s.t.}\ & ||\omega_i-\zeta_j||^L \beta^1 + \beta^2_j  \geq V(x,\omega_i)-\eta& i\in \mathcal{M}, j \in \mathcal{K} \label{wass2-c2} \\
       & ||\omega_i-\zeta_j||^L \beta^1 + \beta^2_j  \geq  0 & i\in \mathcal{M}, j \in \mathcal{K}   \label{wass2-c3}
    \end{align}
\end{subequations}

Replacing the expectation and risk terms from models \eqref{model:wass-expect} and \eqref{model:wass-cvar} in the original formulation gives us

\begin{subequations}
    \begin{align}
        \min \ &\ \CeO \xe + \CgO \xg+\lambda (\mathfrak{D}\beta^1+\sum_{j\in \mathcal{K}} \beta^2_j q_j)+ \notag \\
        & (1-\lambda)\left[\frac{1}{1-\alpha}( \mathfrak{D}\beta^1+\sum_{j\in \mathcal{K}} \beta^2_j q_j) +\eta \right] \notag \\     
      \text{s.t.}\ & \eqref{wass-c1}, \eqref{wass-c2}, \eqref{wass2-c2}, \eqref{wass2-c3}, \eqref{val_func:c1}-\eqref{val_func:c4} \\
      & \xe,\xg \in \mathbb{Z}^+ \times \mathbb{R}^+,\ye_i,\yg_i \in \mathbb{R}^+
    \end{align}
\end{subequations}

\noindent which is equivalent to the following MILP:
\begin{subequations}
    \begin{align}
        \min \ &\ \CeO \xe + \CgO \xg+\lambda (\mathfrak{D}\beta^1+\sum_{j\in \mathcal{K}} \beta^2_j q_j)+ \notag \\
        & (1-\lambda)\left[\frac{1}{1-\alpha}( \mathfrak{D}\beta^1+\sum_{j\in \mathcal{K}} \beta^2_j q_j) +\eta \right] \notag \\   
      \text{s.t.}\ & ||\omega_i-\zeta_j||^L \beta^1 + \beta^2_j  \geq V(x,\omega_i)& i\in \mathcal{M}, j\in \mathcal{K}\\
      & ||\omega_i-\zeta_j||^L \beta^1 + \beta^2_j  \geq V(x,\omega_i)-\eta& i\in \mathcal{M}, j\in \mathcal{K}\\
      & ||\omega_i-\zeta_j||^L \beta^1 + \beta^2_j  \geq  0 & i\in \mathcal{M}, j \in \mathcal{K}\\
      & \BeO\ye_i+\AeO\xe=\de_i & i\in \mathcal{M} \\
& \BeT\ye_i+\Fv_i\AeT\xe=\bT & \nu \in \mathcal{V},i\in \mathcal{M} \\
&\BgO\yg_i+\AgO\xg=\dg_s & i\in \mathcal{M}\\
& \Be\ye_i+\Bg\yg_i+\Ae\xe+\Ag\xg=\bb & i\in \mathcal{M} \\
      &\xe,\xg \in \mathbb{Z}^+ \times \mathbb{R}^+,\ye_i,\yg_i, \beta^1\in \mathbb{R}^+, \beta^2_i,\delta^1_j, \delta^2 \in \mathbb{R}
    \end{align}
\end{subequations}

\section{SP Model} \label{App-sec:SP-model}
Let the uncertain power demand, capacity factors of VREs, and gas demand have a  probability distribution $\mathbb{P}$ with finite and discrete support $\Xi=\{\xi^1, \xi^2,\ldots, \xi^S \}$ where $\mathbb{P}(\xi^s)=p_s$ for all $s\in \mathcal{S}=\{1,2,\ldots, S\}$. 
Let $\de_s$, $\Fv_s$, and $\dg_s$ to denote the power demand, capacity factors of VREs, and gas demand for scenario $s$. Also let $\xeg=(\xe,\xg)$ and $\boldsymbol{\omega}_s =(\de_s,\dg_s,\Fv_s)$ to simplify the exposition.
Then, the two-stage stochastic programming model with a mean-CVaR objective for the joint power-gas planning is formulated as:

\begin{align}\label{model:sp_orig}
\min_{x\in \mathcal{X}} \ & \CeO \xe + \CgO \xg +\lambda \mathbb{E}_{\mathbb{P}}[V(\xeg,\boldsymbol{\omega}_s)]+(1-\lambda)\cvar^\alpha_{\mathbb{P}}(V(\xeg,\boldsymbol{\omega}_s))
\end{align}



\noindent where in the case of stochastic programming with discrete scenarios, the \cvar can equivalently be formulated as:

$$
\cvar^\alpha_{\mathbb{P}}(V(\xeg,\boldsymbol{\omega}_s)) =\min_{\eta\in \mathbb{R}} \{\eta+\frac{1}{1-\alpha}\sum_{s\in \mathcal{S}}p_sV_s: V_s\geq \theta_s-\eta, \quad s\in \mathcal{S}, V_s\in \mathcal{R}^+ \} 
$$

Then the extensive formulation of problem~\eqref{model:sp_orig} becomes:
\begin{subequations}
\begin{align}
\min \ &\CeO \xe+\CgO \xg +\lambda \sum_{s\in \mathcal{S}}p_s\theta_s+(1-\lambda)(\eta+\frac{1}{1-\alpha}\sum_{s\in \mathcal{S}}p_sV_s)\\
\text{s.t. }\ & \AeO\xe+\BeO\ye_s=\de_s & s\in \mathcal{S} \\
& \Fv_s\AeT\xe+\BeT\ye_s=\bT & \nu \in \mathcal{V}, s\in \mathcal{S} \\
&\AgO\xg+\BgO\yg_s=\dg_s & s\in \mathcal{S}\\
& \Ae\xe+\Be\ye_s+\Ag\xg+\Bg\yg_s=\bb & s\in \mathcal{S}\\ 
& \theta_s = \CeT \ye_s+\CgT \yg_s & s\in \mathcal{S}\\
&V_s\geq \theta_s-\eta & s\in \mathcal{S} \\
&\xe,\xg \in \mathbb{Z}^+ \times \mathbb{R}^+,\ye_s,\yg_s,\eta, V_s\in \mathbb{R}^+
\end{align}
\end{subequations}

\section{The Role of Spatial Correlation in the MDRO Model}\label{app:kappa}
We formulate the moment-based DRO (MDRO) in Section~\ref{sec:DRO-model} where its ambiguity sets incorporate spatial correlation and distance between nodes of the same energy vector. Through adjustment factor $\kappa$, these correlations and geographical distances affect the deviation of the first moment from the nominal values. Table~\ref{tab:kappa} shows the sensitivity analysis on the value of $\kappa$ for instances of the RF electrification scenario. As shown, the problem is robust to deviations from the nominal values. Only in large values of $\kappa$ does the objective function value increase for up to 3.6\% in 80\% and 1.9\% in 95\% decarbonization targets. Therefore, the

\begin{table}[htbp]
\centering 
\caption{The sensitivity of total system cost to $\kappa$ values in the MDRO model. The values are the percentage change compared to the total system cost when $\kappa=1$.}
\label{tab:kappa}
\setlength{\tabcolsep}{3pt}
  \renewcommand{\arraystretch}{0.95} 
\begin{tabular}{c|ccccc}
\toprule
$\kappa$& 1e1 & 1e2 & 1e3 & 1e4 & 1e6 \\
\midrule 
80\%& 0.5&0.5&0.5& 1.1 & 3.6 \\
95\%  & 0& 0&0&0.4&1.9 \\
\bottomrule
\end{tabular}
\end{table}

\section{Two-stage Stochastic Programming Model}\label{model-sp}

All proposed methods in the paper including SP, WDRO and MDRO are based on \ModelName which originally is a deterministic model to minimize the planning cost decisions for power and gas systems under the two systems' interdependency. The full description of the \ModelName is given in \citep{KhorramfarEtal2024_AE, Khorramfar2025_CRS}. This section presents the full description of the mathematical model implemented for the SP. Apart from the stochasticity of the supply and demand parameters, in our mathematical models, we turn off some of the features in the \ModelName including import from other regions, carbon capture and storage infrastructure, long-duration storage, planning reserve margin, low-carbon-fuel supply curve, and unit commitment. 

\subsection{Nomenclature} \label{SIsec:nomenclature}
\begin{table}[H]
    \centering
    \begin{tabular}{l|l}
    \textbf{Sets}&\\
    \toprule \midrule 
    $\mathcal{N}^{\text{e}}$, $\mathcal{P}$, $\mathcal{S}$ & Power system nodes, power plant types, scenarios\\
    $\mathcal{R} \subset \mathcal{P}$ & VRE power plant types 
    \\
    $\mathcal{G} \subset \mathcal{P}$ & gas-fired plant types 
    \\
    $\mathcal{CCS} \subset \mathcal{P}$ & gas-fired plant types with carbon capture technology
    \\
    $\mathcal{H} \subset \mathcal{P}$ & Thermal plant types  
    \\
    $\mathcal{Q} \subset \mathcal{P}$ & Technology with a resource availability limit\\
    $\mathcal{Q}'  $ & Set of technologies with resource availability limits \\
    $\mathfrak{R}$, $\mathcal{T}^{\text{e}}$& Representative days, index set of representative hours for  power system\\
    $\mathfrak{T}^{\text{e}}_\tau$, $t^{\text{start}}_\tau,t^{\text{end}}_\tau$& Hours in $\mathcal{T}^{\text{e}}$ that are represented by day $\tau$, First and last hour in $\mathfrak{T}^{\text{e}}_\tau$\\
     $\mathcal{L}^{\text{e}}$, $\mathcal{L}^{\text{e}}_{nm}$ & Existing and candidate transmission lines$\backslash$between node $n$ and $m$\\
    $\mathcal{S}^{\text{e}}_n$, $\mathcal{A}^{\text{g}}_n$ &  All energy storage systems types, adjacent gas  nodes for node  $n$\\
    \hdashline \vspace{-0.3cm}\\
    $ \mathcal{N}^{\text{g}}, \mathcal{N}^s$ & Gas and SVL nodes\\
     $\mathcal{T}^{\text{g}}$, $\mathcal{A}^s_k$& Days of the planning year, adjacent SVL facilities of node  $k$\\
$\mathcal{L}^{\text{g}}$ & Existing and candidate pipelines\\
    $\mathcal{L}^{\text{gExp}}_{k}$, $\mathcal{L}^{\text{gImp}}_{k}$ & Existing and candidate pipelines starting from$\backslash$ endin at node $k$\\
    \hspace{0.15cm}$\mathcal{L}^{\text{LCF}}$ &  LCF availability levels and prices\\
            \bottomrule
    \end{tabular}
\end{table}

\begin{table}[]
    \centering
    \begin{tabular}{l|l}
    \textbf{Indices}&\\
    \toprule \midrule 
 $n,m, k, j, i$ & Power system node, gas system node, SVL facility node, power generation plant type \\
    $r, \ell$ & Storage type for power network, electricity transmission line or gas pipeline\\
    $t/ \tau, l$ & Time step for power/gas system's operational periods\\
            \bottomrule
    \end{tabular}
\end{table}

\begin{table}[]
    \centering
    \begin{tabular}{l|l}
    \textbf{Annualized} &\textbf{Cost Parameters}\\ \toprule \midrule 
    $C^{\text{inv}}_{i}, C^{\text{dec}}_i$ &  CAPEX of plants, plant decommissioning cost, [$\$$/plant] \\
         ${C}^{\text{trans}}_{\ell}$ & Transmission line establishment cost, [$\$$/line] \\
        $C^{\text{EnInv}}_{r}, C^{\text{pInv}}_{r}$ & Storage establishment energy/power-related cost, [$\$$/MWh]\\
        \hdashline \vspace{-0.3cm}\\     
      $C^{\text{pipe}}_{\ell}, C^{\text{pipeDec}}_{\ell}$ &  Pipelines establishment cost, decommissioning cost for pipeline $\ell$ [$\$$/line] \\
      $C^{\text{strInv}}_{j}, C^{\text{vprInv}}_{j}$ &  CAPEX of storage tanks/vaporization plants at SVLs, [$\$$/MMBtu], [$\$$/MMBtu/hour]\\ 
            \bottomrule
    \end{tabular}
\end{table}

\begin{table}[]
    \centering
    \begin{tabular}{l|l}
    \textbf{Annual Costs} &\textbf{}\\ \toprule \midrule 
  $C^{\text{fix}}_{i}, C^{\text{trFix}}_{\ell}$ & FOM for plants and transmission lines, [$\$$], [$\$$/line] \\
        $C^{\text{EnFix}}_{r}, C^{\text{pFix}}_{r}$ & Energy and energy-related FOM for storage, [$\$$/MWh], [$\$$/MW]\\
        \hdashline \vspace{-0.3cm}\\
      $C^{\text{strFix}}_{j}, C^{\text{vprFix}}_{j}, C^{\text{pipeFix}}_{\ell}$ & {\small FOM for storage tanks, vap. plants, and pipelines, [$\$$/MMBtu], [$\$$/MMBtu/hour], [$\$$/line]}\\
            \bottomrule
    \end{tabular}
\end{table}

\begin{table}[]
    \centering
    \begin{tabular}{l|l}
    \textbf{Other Cost} &\textbf{Parameters}\\ \toprule \midrule 
$C^{\text{var}}_{i}, C^{\text{startUp}}_i$ & VOM for  plants, start-up cost for plants [$\$$/MWh], [$\$$]\\
     $C^{\text{eShed}}, C^{\text{fuel}}_i$ & Unsatisfied power demand cost,  fuel price for plants [$\$$/MWh], [$\$$/MMBtu] \\
     \hdashline \vspace{-0.3cm}\\
     $C^{\text{ng}}, C^{\text{gShed}}$ & Fuel price for NG, unsatisfied gas demand cost, [$\$$/MMBtu]\\
            \bottomrule
    \end{tabular}
\end{table}

\begin{table}[]
    \centering
    \begin{tabular}{l|l}
    \textbf{Other Power} &\textbf{System Parameters}\\ \toprule \midrule 
   $\rho_{ntis}, U^{\text{prod}}_{i}$ & Capacity factor for renewable plants, nameplate capacity, [MW]\\
         $D^{\text{e}}_{nts}, h_i$ & Power demand, heat rate, [MWh], [MMBtu/MWh]\\
       $b_{\ell}, \eta_i$ & Susceptance of line $\ell\in \mathcal{L}^{\text{e}}$, carbon capture rate, [\%]\\
        $L^{\text{prod}}_{i}, U^{\text{ramp}}_{i}$ & Minimum stable output, ramping limit [$\%$]\\
        $\gamma^{\text{eCh}}_r, \gamma^{\text{eDis}}_r, \gamma^{\text{loss}}_r$ & Storage charge, discharge, and hourly self-discharge rate\\
        $I^{\text{trans}}_{\ell}, U^{\text{trans}}_\ell$ & Initial/upper bound capacity for transmission line $\ell$, [MW]\\
        $\mathcal{I}^{\text{trans}}_{\ell}$ & 1, if  trans. line $\ell$ exists; 0, otherwise\\
        ${I}^{\text{num}}_{ni}$ & Initial number of plants\\
        $U^{\text{e}}_{\text{emis}}, \zeta$ & Baseline emiss. of CO$_2$ in 1990 from gen. consumption [ton], emissions reduction goal\\
        $U^{\text{prod}}_{\mathcal{Q}}$ & Production capacity for set of plants $\mathcal{Q}\subset \mathcal{P}$, [MW]\\
        $w_{ts}, \phi^{\text{e}}_{ts}$ & Weight of the rep. period $t$, mapping of rep. period $t$ to its orig. period in the time series\\
        $R^{\text{CRM}}, \gamma^{\text{CRM}}_{nit}$ & Capacity reserve margin rate, cap. derating factor of plant type $i$ at node $n$ at time $t$\\
            \bottomrule
    \end{tabular}
\end{table}

\begin{table}[]
    \centering
    \begin{tabular}{l|l}
    \textbf{Other Gas} &\textbf{System Parameters}\\ \toprule \midrule 
 $D^{\text{g}}_{k\tau s}, \eta^{\text{g}}$ &  Gas demand [MMBtu], emission factor for NG [ton CO$_2$/MMBtu]\\
        $U^{\text{inj}}_k, I^{\text{gStr}}_{j}$ & Upper bound for gas supply, initial storage capacity,  [MMBtu]\\
        $\gamma^{\text{liqCh}}_j, \gamma^{\text{vprDis}}_j$ & Charge efficiency of liquefaction plant, discharge efficiency of vaporization plant\\
        $\beta, I^{\text{store}}_{kj}$ & Boil-off gas coefficient,  initial capacity of storage facility\\
        $I^{\text{pipe}}_{\ell}, U^{\text{pipe}}_{\ell}$ & Initial capacity for pipeline $\ell$, upper bound capacity for pipeline $\ell$, [MMBtu/day]\\
        $\mathcal{I}^{\text{pipe}}_{\ell}, \Omega_n$ & 1, if the pipeline $\ell$ exists; 0, otherwise, representative day for day $n$\\
        $I^{\text{vpr}}_{j}, I^{\text{liq}}_{j}$ & Initial vaporization and liquefaction capacity, [MMBtu/d]\\
        $U^{\text{g}}_{\text{emis}}$ & Baseline emission of CO$_2$ in 1990 from non-generation consumption, [ton]\\
            \bottomrule
    \end{tabular}
\end{table}

\begin{table}[]
    \centering
    \begin{tabular}{l|l}
    \textbf{Investment} &\textbf{Decision Variables}\\ \toprule \midrule 
 $x^{\text{op}}_{ni}\in \mathbb{Z}^+, x^{\text{op}}_{ni}\in \mathbb{R}^+$ & Number of available thermal and VRE plants\\          
         $x^{\text{est}}_{ni}\in \mathbb{Z}^+, x^{\text{est}}_{ni}\in \mathbb{R}^+$ & Number of new thermal and VRE plants established \\
         $x^{\text{dec}}_{ni}\in \mathbb{Z}^+, x^{\text{dec}}_{ni}\in \mathbb{R}^+$ & Number decommissioned thermal and vRE plants \\
         $y^{\text{eCD}}_{nr}\in \mathbb{R}^+, y^{\text{eLev}}_{nr}\in \mathbb{R}^+$& Charge/discharge capacity of storage battery, battery storage level\\
         $z^{\text{eInv}}_\ell\in \mathbb{B}$& 1, if transmission line $\ell$ is built; 0, otherwise\\
         $z^{\text{gInv}}_\ell, z^{\text{gDec}}_\ell, z^{\text{gOp}}_\ell\in \mathbb{B}$& 1, if pipeline $\ell$ is built/decommissioned/operational; 0, otherwise\\
            \bottomrule
    \end{tabular}
\end{table}

\begin{table}[]
    \centering
    \begin{tabular}{l|l}
    \textbf{Other} &\textbf{Power System Decision Variables}\\ \toprule \midrule 
$p_{ntis}, f^{\text{e}}_{\ell ts}\in \mathbb{R}^+$& Generation and flow rate, [MW]\\
        $s^{\text{eCh}}_{ntrs}, s^{\text{eDis}}_{ntr}, s^{\text{eLev}}_{ntr}\in \mathbb{R}^+$& Storage charged/discharged/level, [MW] [MWh] \\
         $a^{\text{e}}_{nts}, \mathcal{E}^{\text{e}}_s\in \mathbb{R}^+$ & Amount of load shedding [MWh], total emission from power system\\
            \bottomrule
    \end{tabular}
\end{table}

\begin{table}[]
    \centering
    \begin{tabular}{l|l}
    \textbf{Other} &\textbf{Gas System Decision Variables}\\ \toprule \midrule 
$x^{\text{gStr}}_{j}, x^{\text{vpr}}_{j}\in \mathbb{R}^+$ & Installed additional  storage/vaporization capacities\\
        $f^{\text{g}}_{\ell\tau s},f^{\text{ge}}_{kn\tau s}, f^{\text{gl}}_{kj\tau s} \in {\mathbb{R}}^+$& Flow rates/from gas nodes to power nodes/from gas nodes to  liquefaction plants\\
        $f^{\text{vg}}_{jk\tau s}, g_{k\tau s} \in {\mathbb{R}^+}$& Flow rates from vaporization plants to gas nodes, gas supply (injection)\\        
        $s^{\text{gStr}}_{j\tau s}, s^{\text{vpr}}_{j\tau s}, s^{\text{liq}}_{j\tau s} \in \mathbb{R}^+$&  Storage capacities, vaporization and liquefaction amounts\\
        $a^{\text{g}}_{k\tau s}, a^{\text{LCF}}_{k\tau ls}\in \mathbb{R}^+$ & Amount of load shedding and LCF consumed\\
        $\mathcal{E}^{\text{g}}_s$ & Total emission from gas system\\
            \bottomrule
    \end{tabular}
\end{table}

\subsection{Mathematical Model}
The SP model's objective function is the summation of model~\eqref{elec-obj} and \eqref{ng-obj}. The constrains of SP consist of constraints presented in models~\eqref{pow-constraints}, \eqref{gas-constraints}, \eqref{coupling-costraints}.

\noindent\textbf{Power System Objective Function:}
\begin{subequations} \label{elec-obj}
\begin{align}
      \min & \sum_{n \in \mathcal{N}^{\text{e}}}\left[ \sum_{i \in \mathcal{P}}  (C^{\text{inv}}_i x^{\text{est}}_{ni}+C^{\text{fix}}_{i} x^{\text{op}}_{ni} +C^{\text{dec}}_i x^{\text{dec}}_{ni}) + \sum_{r \in \mathcal{S}^{\text{e}}_n}\left((C^{\text{pInv}}_r+C^{\text{pFix}}) y^{\text{eCD}}_{nr} +(C^{\text{EnInv}}_r+C^{\text{EnFix}}) y^{\text{eLev}}_{nr}\right)\right] \notag\\
      &+ \sum_{l\in \mathcal{L}^{\text{e}}}  C^{\text{trans}}_{\ell} z^{\text{eInv}}_\ell \notag +\sum_{\ell \in \mathcal{L}^{\text{e}}:I^{\text{trans}}_\ell=1} C^{\text{trFix}}_\ell +\sum_{\ell \in \mathcal{L}^{\text{e}}:I^{\text{trans}}_\ell=0} C^{\text{trFix}}_\ell z^{\text{eInv}}_\ell  \\  
         &+  \sum_{n \in \mathcal{N}^{\text{e}}}\sum_{t\in \mathcal{T}^{\text{e}}} \left[ w_{ts} C^{\text{eShed}}_n a^{\text{e}}_{nts} + \sum_{i \in \mathcal{P}}w_{ts} p_{ntis} C^{\text{fuel}}_{i} h_i + w_{ts} p_{ntis}C^{\text{var}}_{i}\right]
\end{align}
\end{subequations}

The objective function~\eqref{elec-obj} minimizes the total investment and operating costs incurred in the power system including investment in new power plants, fixed operation and maintenance (FOM) of power plants, decommission of power plants, investment in storage, investment in transmission lines, FOM of transmission lines, load shedding, fuel, and  variable operating and maintenance  (VOM) cost of operating power plants.

\vspace{0.5cm}
\noindent\textbf{Power System Constraints:}
For every $n\in \mathcal{N}^{\text{e}},i\in \mathcal{P}$ $l\in \mathcal{L}^{\text{e}},t \in \mathcal{T}^{\text{e}},  n,m\in \mathcal{N}^{\text{e}}_\ell, r\in \mathcal{S}^{\text{e}}_n, s\in \mathcal{S}$ the constraints are defined as:
\begin{subequations}\label{pow-constraints}
\begin{align}
      & x^{\text{op}}_{ni} = I^{\text{num}}_{ni}-x^{\text{dec}}_{ni}+x^{\text{est}}_{ni} &  \label{elec-c1}\\
        &L^{\text{prod}}_{i} x_{ni} \leq  p_{ntis} \leq   U^{\text{prod}}_{i}x_{ni}&\hspace{-2cm} i\in \mathcal{H} 
    \label{elec-c4}\\
    &| p_{ntis} -p_{n,(t-1),i}| \leq U^{\text{ramp}}_{i}U^{\text{prod}}_{i}x^{ni}   & \hspace{-2cm} i\in \mathcal{H} \label{elec-c5}\\
     &p_{ntis} \leq \rho_{ntis}U^{\text{prod}}_{i} x^{\text{op}}_{ni} &\hspace{-2cm} i \in \mathcal{R}  \label{elec-c6}\\
     & a^{\text{e}}_{nts}\leq D^{\text{e}}_{n\phi^{\text{e}}_{ts}}& \label{elec-c7} \\
      &\sum_{i \in \mathcal{P}}p_{ntis} +\sum_{m\in \mathcal{N}^{\text{e}}}\sum_{l\in \mathcal{L}^{\text{e}}_{nm}}\sign(n-m) f^{\text{e}}_{\ell ts}+\sum_{r\in \mathcal{S}^{\text{e}}_n} (s^{\text{eDis}}_{ntr}-s^{\text{eCh}}_{ntrs})+ a^{\text{e}}_{nts}=D^{\text{e}}_{n \phi^{\text{e}}_{ts}}       \label{elec-c8}   \\
    & \vert f^{\text{e}}_{lt }\vert \leq I^{\text{trans}}_{\ell}&\hspace{-2cm}  \text{if } \mathcal{I}^{\text{trans}}_\ell=1   \label{elec-c9}\\
      & \vert f^{\text{e}}_{lt }\vert \leq U^{\text{trans}}_\ell z^{\text{eInv}}_{\ell}& \hspace{-2cm}  \text{if } \mathcal{I}^{\text{trans}}_\ell=0  \label{elec-c10}\\
      & s^{\text{eLev}}_{n t r}=(1-\gamma^{\text{loss}}_{r}) (s^{\text{eLev}}_{ntr})+ \gamma^{\text{eCh}}_r s^{\text{eCh}}_{ntr}-\frac{s^{\text{eDis}}_{n t r}}{\gamma^{\text{eDis}}_r}& \hspace{-4cm} t\in \mathcal{T}^{\text{e}}\backslash \{t^{\text{start}}_\tau \lvert \  \tau \in \mathfrak{R} \}   \label{elec-c16}\\
       & s^{\text{eLev}}_{n 0 r}= \gamma^{\text{eCh}}_r s^{\text{eCh}}_{n0 r}-\frac{s^{\text{eDis}}_{n 0 r}}{\gamma^{\text{eDis}}_r}&   \label{elec-c15}\\
    & s^{\text{eCh}}_{ntrs}\leq y^{\text{eCD}}_{nr}, \quad s^{\text{eCh}}_{ntrs}\leq y^{\text{eCD}}_{nr}, \quad s^{\text{eLev}}_{ntr}\leq y^{\text{eLev}}_{nr} &  \label{elec-c21}\\
     & \sum_{n\in \mathcal{N}^{\text{e}}}\sum_{i\in \mathcal{Q}}  U^{\text{prod}}_i x^{\text{op}}_{ni} \leq U^{\text{prod}}_{\mathcal{Q}} & \mathcal{Q}\in \mathcal{Q}'
   \label{elec-c24}
\end{align}
\end{subequations}

Constraints~\eqref{elec-c1} model the number of operating plants. The generation limits are imposed in constraints~\eqref{elec-c4}. 
Constraints~\eqref{elec-c5} are the ramping constraints. The generation pattern of VREs is imposed in constraints~\eqref{elec-c6}. Constraints~\eqref{elec-c7} state that the load-shedding amount can not exceed demand. Constraints~\eqref{elec-c8} ensure the power balance. Flow for the existing transmission lines is limited by constraints~\eqref{elec-c9}. Constraints~\eqref{elec-c10} limit the flow in candidate transmission lines only if it is already established (i.e., $z^{\text{t}}_{\ell}$=1). 
Constraints~\eqref{elec-c16}-\eqref{elec-c21} model battery storage dynamics and storage limit. Finally, constraint~\eqref{elec-c24} limits the installed capacity of a certain set of power plants to their maximum availability limit.

\noindent\textbf{Gas System Objective Function:} 
\begin{subequations}\label{ng-obj}
\begin{align}
    \min & \sum_{l\in \mathcal{L}^{\text{g}}} \left (C^{\text{pipe}}_{\ell} z_{\ell}^{\text{g}} + c^{\text{pipeDec}}_\ell z^{\text{gDec}}\ell + c^{\text{pipeFix}}_\ell z^{\text{gOp}}_\ell \right)  \\
    &+ \sum_{k \in \mathcal{N}^{\text{g}}} \sum_{\tau \in \mathcal{T}^{\text{g}}}  C^{\text{ng}} g_{k\tau} + C^{\text{gShed}} a^{\text{ng}}_{k\tau }\\
    &+ \sum_{j \in \mathcal{N}^s}(C^{\text{strInv}}_j x^{\text{gStr}}_{j} +C^{\text{vprInv}}_j x^{\text{vpr}}_{j}) +C^{\text{strFix}}(I^{\text{gStr}}_{j}+x^{\text{gStr}}_j)+C^{\text{vprFix}}(I^{\text{vpr}}_{j}+x^{\text{vpr}}_j)\\
\end{align}
\end{subequations}

The objective function~\eqref{ng-obj} minimizes the total investment and operating costs incurred in the gas system including investment/decommissioning/FOM of pipelines, import, load shedding, and storage/vaporization/liquefaction costs. 

\vspace{0.5cm}
\noindent\textbf{Gas System Constraints:} 
For every $k\in \mathcal{N}^{\text{g}}, \tau \in \mathcal{T}^{\text{g}} $ $\ell \in \mathcal{L}^{\text{g}}, j\in \mathcal{N}^s, s\in \mathcal{S}$

\begin{subequations}\label{gas-constraints}
\begin{align}
   &g_{k\tau} -\sum_{l \in \mathcal{L}^{\text{gExp}}_{k}} f^{\text{g}}_{\ell\tau s}+\sum_{l \in \mathcal{L}^{\text{gImp}}_{k}} f^{\text{g}}_{\ell\tau s}-\sum_{n\in \mathcal{A}^{\text{e}}_k} f^{\text{ge}}_{kn\tau } +\sum_{j\in \mathcal{A}^s_k} (f^{\text{vg}}_{jk \tau}-f^{\text{gl}}_{kj\tau s})+ a^{\text{LCF}}_{k\tau}+a^{\text{g}}_{k\tau s}=D^{\text{g}}_{k\tau s} &  \label{ng-c1}\\
& L^{\text{inj}}_k \leq g_{k\tau}+a^{\text{LCF}}_{k\tau}\leq U^{\text{inj}}_k & \label{ng-c2}\\
& f^{\text{g}}_{\ell \tau } \leq U^{\text{pipe}}_{\ell} z^{\text{gOp}}_{\ell}&  \label{ng-c6}\\
    &\sum_{k \in \mathcal{N}^{\text{g}}:j\in \mathcal{A}^s_k} f^{\text{gl}}_{kj \tau} =s^{\text{liq}}_{j\tau s} & \label{ng-c7}\\
    &\sum_{k \in \mathcal{N}^{\text{g}}:j\in \mathcal{A}^s_k } f^{\text{vg}}_{jk \tau}=s^{\text{vpr}}_{j\tau s}  & \label{ng-c8}\\
     &s^{\text{gStr}}_{j\tau s} = (1-\beta) s^{\text{gStr}}_{j,\tau-1}+\gamma^{\text{liqCh}}_j s^{\text{liq}}_{j\tau s}-\frac{s^{\text{vpr}}_{j\tau s}}{\gamma^{\text{vprDis}}_j}  &  \label{ng-c9}\\
    &s^{\text{vpr}}_{j\tau s}\leq  I^{\text{vpr}}_{j}+x^{\text{vpr}}_j  &  \label{ng-c10}\\
    &s^{\text{gStr}}_{j\tau s}\leq  I^{\text{gStr}}_{j}+x^{\text{gStr}}_j  &  \label{ng-c11}\\
    &z^{\text{gOp}}_\ell =  \mathcal{I}^{\text{pipe}} + z^{\text{gInv}}_\ell - z^{\text{gDec}}_\ell  &  \label{ng-c12}
\end{align}
\end{subequations}
Constraints~\eqref{ng-c1} ensure the gas balance.
The gas fuel import limits are imposed in constraints~\eqref{ng-c2}. 
The constraints~\eqref{ng-c6}  limit the flow between gas nodes for operational pipelines, respectively. The flow to liquefaction facilities is calculated in constraints~\eqref{ng-c7}. Similarly, the flow out of vaporization facilities is modeled via constraints~\eqref{ng-c8}.
Constraints~\eqref{ng-c9} ensure the storage balance. Constraints~\eqref{ng-c10} and \eqref{ng-c11} limit the capacity of vaporization and storage tanks to their initial capacity plus the increased capacity, respectively.
Pipeline $\ell$ is operational if either it is existing and not decommissioned, or newly established. 

\vspace{0.5cm}

\noindent \textbf{Coupling Constraints}
The following constraints are coupling constraints that relate operational decisions of the power and gas systems together.
\begin{subequations}\label{coupling-costraints}
\begin{align}
    & \sum_{k \in \mathcal{A}^{\text{e}}_n} f^{\text{ge}}_{k n\tau } =   \sum_{t \in \mathfrak{T}^{\text{e}}_\tau} \sum_{i \in \mathcal{G}} h_i p_{ntis} &\hspace{-2cm} n\in \mathcal{N}^{\text{e}}, \tau \in \mathfrak{R}\label{coup-1}\\
 &\mathcal{E}^{\text{e}}_s = \sum_{n\in \mathcal{N}^{\text{e}}}\sum_{t\in \mathcal{T}^e}\sum_{i \in \mathcal{G}} w_{ts}(1-\eta_i)\eta^{\text{g}} h_i p_{ntis}&\notag \\
 &\mathcal{E}^{\text{g}}_s =\sum_{k \in \mathcal{N}^g}\sum_{\tau \in \mathcal{T}^g} \eta^{\text{g}}( D^{\text{g}}_{k\tau s}-  a^{\text{LCF}}_{k\tau}-a^{\text{g}}_{k\tau s})\notag \\
 &  \mathcal{E}^{\text{e}}_s+\mathcal{E}^{\text{g}}_s \leq (1-\zeta) (U^{\text{e}}_{\text{emis}} +U^{\text{g}}_{\text{emis}}) &\label{coup-2}
\end{align}
\end{subequations}
The first coupling constraints~\eqref{coup-1} capture the flow of gas to the power network for each node and at each time period. 
The variable $\mathcal{E}^{\text{e}}_s$ accounts for CO$_2$ emission due to the consumption of gas in the power system. The variable  $\mathcal{E}^{\text{g}}_s$ computes the emission from the gas system by subtracting the demand from LCF consumption and gas load shedding. The second coupling constraint~\eqref{coup-2} ensures that the net CO$_2$ emissions associated with the power-gas system are below a pre-specified threshold value, which is defined based on a baseline (e.g., historical) emissions level. The first term is the emissions due to non-power gas consumption (i.e., gas consumption in the gas system such as space heating, industry use, and transportation). Since the model does not track whether LCF is used to meet non-power gas demand or for power generation, the first term computes gross emissions from all gas use presuming it is all fossil and then subtracts emissions benefits from using LCF.

\section{Sequential Construction Method for the Stochastic Program} \label{SCM4SP}
This section details the Sequential Construction Method (SCM) for the model presented in section~\eqref{model-sp}. 

\begin{enumerate}
\item Solve {\textit{MILP}$^{\text{ref}}$} with the following simplifications and get the solution $\bar{X}$:
\begin{itemize}
\item Remove all network constraints for the power system and get the \textit{copper-plate model}. In effect, the constrains (2g) and (2h) are removed from the model presented in section~\eqref{model-sp}, and the constraint (2f) becomes the following for all $t\in \mathcal{T}^{\text{e}}$
$$
\sum_{n\in \mathcal{N}^{\text{e}}}\left[\sum_{i \in \mathcal{P}}p_{ntis} +\sum_{r\in \mathcal{S}^{\text{e}}_n} (s^{\text{eDis}}_{ntr}-s^{\text{eCh}}_{ntrs})+ a^{\text{e}}_{nts}\right]=\sum_{n\in \mathcal{N}^{\text{e}}} D^{\text{e}}_{n \phi^{\text{e}}_{ts}}
$$

\item Relax the integrality of integer variables $x^{\text{op}}_{ni}, x^{\text{est}}_{ni}, x^{\text{dec}}_{ni}, z^{\text{elnv}}_\ell, z^{\text{glnv}}_\ell, z^{\text{gDec}}_\ell, z^{\text{gOp}}_\ell$
\end{itemize}
\item Solve {\textit{MILP}$^{\text{ref}}$} with the following simplifications and constraints and get the solution $\hat{X}$:
\begin{itemize}
\item Relax the integrality of integer variables $x^{\text{op}}_{ni}, x^{\text{est}}_{ni}, x^{\text{dec}}_{ni}, z^{\text{elnv}}_\ell, z^{\text{glnv}}_\ell, z^{\text{gDec}}_\ell, z^{\text{gOp}}_\ell$
\item For thermal plants, set $x^{\text{op}}_{ni} = \lfloor(\bar{x}^{\text{op}}_{ni}) \rceil, n\in\mathcal{N}, i\in \mathcal{H}$, where $\lfloor x \rceil$ is the rounding operator that returns the nearest integer value to $x$, and $\bar{x}^{\text{op}}_{ni}$ is the value of the variable in previous step
\item For VRE plants, add constraints $x^{\text{op}}_{ni} = \lfloor\bar{x}^{\text{op}}_{ni} \rfloor, n\in\mathcal{N}, i\in \mathcal{V}$, where $\lfloor x \rfloor$ is the floor operator that returns the greatest integer less than or equal to $x$, and $\bar{x}^{\text{op}}_{ni}$ is the value of the variable in previous step.
\item Add constraints $z^{\text{gOp}}_\ell=1$ if $\bar{z}^{\text{gOp}}_\ell \geq \epsilon_1$, and $z^{\text{gOp}}_\ell=0$, otherwise.  $\bar{z}^{\text{gOp}}_\ell$ is the value of the variable in previous step
\end{itemize}
\item Solve {\textit{MILP}$^{\text{ref}}$} with the following simplifications and constraints and get feasible solution ${{X}}^*$:
\begin{itemize}
\item Set $x^{\text{op}}_{ni} = \hat{x}^{\text{op}}_{ni}, n\in\mathcal{N}, i\in \mathcal{H}$, where $\hat{x}^{\text{op}}_{ni} $ is the value of the variable in previous step.
\item Set $z^{\text{gOp}}_\ell =  \hat{z}^{\text{gOp}}_\ell$, where $\hat{z}^{\text{gOp}}_\ell$  is the value of the variable in previous step.
\item Add constraints $z^{\text{eInv}}_\ell=1$ if $\hat{z}^{\text{eInv}}_\ell > \epsilon_2$, and $z^{\text{eInv}}_\ell=0$, otherwise. $\hat{z}^{\text{eInv}}_\ell$ is the value of the variable in previous step.
\end{itemize}
\end{enumerate}
In our experiments, $\epsilon_1=0.01 $ and $\epsilon_2=0.3$.

\bibliographystyle{apalike}
\bibliography{Bibliography.bib}

\end{document}